\documentclass[fleqn,usenatbib]{mnras}

\usepackage{newtxtext,newtxmath}

\usepackage[T1]{fontenc}

\usepackage{graphicx}	
\usepackage{subcaption}

\usepackage{hyperref}
\usepackage{color}
\usepackage{soul}
\usepackage{natbib}
\usepackage{hyperref}
\usepackage{amsmath,bm}
\usepackage{xspace}
\usepackage{enumitem}
\usepackage{xifthen}
\usepackage{hyperref}
\usepackage[normalem]{ulem}
\usepackage{comment}

\hypersetup{    
  colorlinks      = {true},
  linkcolor       = {blue},
  citecolor       = {blue},
  urlcolor        = {blue},
}

\graphicspath{{./figs}}

\definecolor{orange}{rgb}{1.0,0.5,0.}

\def\MDM{\ifmmode{\>M_{\textnormal{\sc dm}}}\else{$$M_{\textnormal{\sc dm}}}\fi}

\def\XH{\ifmmode{\>X_{\textnormal{\sc h}}} \else{$X_{\textnormal{\sc h}}$}\fi}
\def\nH{\ifmmode{\>n_{\textnormal{\sc h}}} \else{$n_{\textnormal{\sc h}}$}\fi}

\def\maspyr{\ifmmode{\>\textnormal{mas~yr}^{-1}}\else{mas~yr$^{-1}$}\fi}

\def\mG{\ifmmode{\>\mu\mathrm{G}}\else{$\mu$G}\fi}
\def\erg{\ifmmode{\> {\rm erg}}\else{erg}\fi}
\def\keV{\ifmmode{\> {\rm keV}}\else{keV}\fi}

\def\deg{\ifmmode{\>^{\circ}}\else{$^{\circ}$}\fi}
\def\onedeg{\ifmmode{\>1^{\circ}}\else{$1^{\circ}$}\fi}

\def\xvir{\ifmmode{\>\!x_{vir}}\else{$x_{vir}$}\fi}
\def\Mvir{\ifmmode{\>\!M_{vir} }\else{$M_{vir} $}\fi}
\def\rvir{\ifmmode{\>\!r_{vir}}\else{$r_{vir}$}\fi}
\def\vvir{\ifmmode{\>\!v_{vir}}\else{$v_{vir}$}\fi}
\def\Vvir{\ifmmode{\>\!V_{vir} }\else{$V_{vir} $}\fi}

\def\tratio{\ifmmode{\>\tau}\else{$\tau$}\fi}

\def\rms{\ifmmode{\>r_{\textnormal{\sc ms}}}\else{$r_{\textnormal{\sc ms}}$}\fi}

\def\Mpc{\ifmmode{\>\!{\rm Mpc}} \else{Mpc}\fi}
\def\kpc{\ifmmode{\>\!{\rm kpc}} \else{kpc}\fi}
\def\pkpc{\ifmmode{\>\!{\rm kpc}^{-1}} \else{kpc$^{-1}$}\fi}
\def\pc{\ifmmode{\>\!{\rm pc}} \else{pc}\fi}

\def\Gyr{\ifmmode{\>\!{\rm Gyr}} \else{Gyr}\fi}
\def\Myr{\ifmmode{\>\!{\rm Myr}} \else{Myr}\fi}
\def\yr{\ifmmode{\>\!{\rm yr}} \else{yr}\fi}
\def\pyr{\ifmmode{\>\!{\rm yr}^{-1}}\else{yr$^{-1}$} \fi}
\def\s{\ifmmode{\>\!{\rm s}}\else{s}\fi}
\def\ps{\ifmmode{\>\!{\rm s}^{-1}}\else{s$^{-1}$}\fi}
\def\Hz{\ifmmode{\>\!{\rm Hz}}\else{Hz}\fi}

\def\kms{\ifmmode{\>\!{\rm km\,s}^{-1}}\else{km~s$^{-1}$}\fi}

\def\K{\ifmmode{\>\!{\rm K}}\else{K}\fi}

\def\sr{\ifmmode{\>\!{\rm sr}}\else{sr}\fi}
\def\psr{\ifmmode{\>\!{\rm sr}^{-1}}\else{sr$^{-1}$}\fi}
\def\arcs{\ifmmode{\>\!{\rm arcsec}}\else{arcsec}\fi}
\def\parcs{\ifmmode{\>\!{\rm arcsec}^{-1}}\else{arcsec${-1}$}\fi}
\def\parcss{\ifmmode{\>\!{\rm arcsec}^{-2}}\else{arcsec${-2}$}\fi}

\def\cm{\ifmmode{\>\!{\rm cm}}\else{cm}\fi}
\def\cc{\ifmmode{\>\!{\rm cm}^{3}}\else{cm$^{3}$}\fi}
\def\sqc{\ifmmode{\>\!{\rm cm}^{2}}\else{cm$^{2}$}\fi}
\def\pcc{\ifmmode{\>\!{\rm cm}^{-3}}\else{cm$^{-3}$}\fi}
\def\psc{\ifmmode{\>\!{\rm cm}^{-2}}\else{cm$^{-2}$}\fi}

\def\g{\ifmmode{\>\!{\rm g}}\else{g}\fi}
\def\Msun{\ifmmode{\>\!{\rm M}_{\odot}}\else{M$_{\odot}$}\fi}
\def\hMsun{\ifmmode{\> h^{-1}{\rm M}_{\odot}}\else{$h^{-1}$M$_{\odot}$}\fi}

\def\Zsun{\ifmmode{\>\!{\rm Z}_{\odot}}\else{Z$_{\odot}$}\fi}

\def\Lsun{\ifmmode{\>\!{\rm L}_{\odot}}\else{L$_{\odot}$}\fi}

\def\rayl{\ifmmode{\>\!{\rm R}}\else{R}\fi}
\def\mR{\ifmmode{\>\!{\rm mR}}\else{mR}\fi}

\def\lya{\ifmmode{\>\!{\rm Ly}\alpha}\else{Ly$\alpha$}\fi}

\def\Ha{\ifmmode{\>\!{\rm H}\alpha}\else{H$\alpha$}\fi}
\def\Hb{\ifmmode{\>\!{\rm H}\beta}\else{H$\beta$}\fi}

\def\HI{\ifmmode{\> \textnormal{\ion{H}{i}}} \else{\ion{H}{i}}\fi}
\def\HII{\ifmmode{\> \textnormal{\ion{H}{ii}}} \else{\ion{H}{ii}}\fi}
\def\CIV{\ifmmode{\> \textnormal{\ion{C}{iv}}} \else{\ion{C}{iv}}\fi}
\def\SiIV{\ifmmode{\> \textnormal{\ion{S}{iv}}} \else{\ion{Si}{iv}}\fi}

\def\NH{\ifmmode{\> {\rm N}_{\rm H}} \else{N$_{\rm H}$}\fi}
\def\Ng{\ifmmode{\> {\rm N}_{\rm gas}} \else{N$_{\rm gas}$}\fi}
\def\NHI{\ifmmode{\> {\rm N}_{\HI}} \else{N$_{\HI}$}\fi}
\def\MHI{\ifmmode{\> {\rm M}_{ \HI}} \else{M$_{\HI}$}\fi}

\def\mua{\ifmmode{\>\mu_{ \textnormal{\Ha}}}\else{$\mu_{ \textnormal{\Ha}}$}\fi}
\def\alphabha{\ifmmode{\>\alpha_{B}^{(\textnormal{\Ha})}}\else{$\alpha_{B}^{(\textnormal{\Ha})}$}\fi}

\newcommand{\ramses}{{\sc Ramses}}
\newcommand\agama{{\sc agama}}

\definecolor{newtext}{rgb}{0.00,0.45,0.15}
\definecolor{oldtext}{rgb}{0.75,0.00,0.00}
\definecolor{todocol}{rgb}{0.90,0.45,0.00}

\title[Massive stellar bars at high redshift]{Turbulent gas-rich discs at high redshift: the origin of early massive stellar bars}

\author[Joss Bland-Hawthorn et al.]{
Joss Bland-Hawthorn,$^{1}$\thanks{E-mail: jbh@physics.usyd.edu.au}
Thor Tepper-Garcia,$^{1}$
Chris Hamilton,$^{2}$
Johan M. Villa-Alatorre,$^{1}$
Oscar Agertz,$^{3}$ \newauthor
Takafumi Tsukui$^{4}$ \& Ken Freeman$^{5}$
\\
$^{1}$Sydney Institute for Astronomy, School of Physics, A28, The University of Sydney, NSW 2006, Australia\\
$^{2}$ Institute for Advanced Study, Princeton University, Princeton, USA \\    
$^{3}$Lund Observatory, Division of Astrophysics, Department of Physics, Lund University, Box 118, SE-221 00 Lund, Sweden\\
$^{4}$Kavli Institute, IPMU Building, University of Tokyo, Japan\\
$^{5}$Research School of Astronomy and Astrophysics, Australian National University, Canberra, ACT 2611, Australia
}

\date{Accepted XXX. Received YYY; in original form ZZZ}

\pubyear{\the\year{2026}}

\begin{document}
\label{firstpage}
\pagerange{\pageref{firstpage}--\pageref{lastpage}}
\maketitle

\begin{abstract}
Recent observations combining the power of ALMA and JWST {\it inter alia} have revealed large ($3-7$ kpc), massive ($3-10\times10^{10}\,\mathrm{M}_\odot$) stellar bars at $z=4-5$ when the Universe was only 1.2-1.6 Gyr old. At this early epoch, the host galaxy was baryon-dominated (typically 75\% gas, 25\% stars) within the observed extent of the disc ($8-15$ kpc). Using \textsc{Nexus} $N$-body/hydrodynamic simulations, we show that such bars can form promptly (400$-$800 Myr), provided the disc mass fraction is high ($f_{\rm disc}\gtrsim 70\%$) and the bar is gas-dominated at the time of its formation, consistent with the observations. In this limit, gas-free bars are unstable to vertical bending modes, but a dominant gas component suppresses this instability. Unlike massive bars in the local Universe, these early bars were sites of vigorous star formation, as we show. Remarkably, for gas-rich models with $f_{\rm gas}\lesssim60\%$, the bars develop X-shaped boxy bulges; at higher gas fractions ($f_{\rm gas}> 60\%$), diffusion suppresses resonant orbit trapping and the emerging bar collapses within 1 Gyr to form a classical bulge. The bar formation time, length, mass, and $m=2$ Fourier amplitude are all inversely related to $f_{\rm gas}$. We present a simple analytic model for how stochastic forcing shifts the bar onset time, defined as the time at which the growing bar amplitude reaches a specified threshold.
\end{abstract}




\section{Introduction}
\label{s:intro}

The formation of galactic discs has been one of the fastest developing areas of galactic astronomy in recent years \citep[q.v.][]{guo23,kartaltepe23,sme25}. In particular, there are now a dozen spatially resolved disc galaxies beyond a redshift $z=4$ at a time when the Universe was $1-1.5$ Gyr old \citep{riz22w,sma23,pop23,roman23,tsu24,xia24,boo25,jai25} 
This was an active era when disc galaxies were very gas-rich \citep{tac18} with gas fractions as high as $f_{\rm gas}\approx 80\%$ (defined in Eq. 2).
To remain supported against gravitational collapse, these discs required high star formation rates, consistent with the elevated gas surface densities inferred at those epochs \citep{tac20}. Stellar feedback, coupling to the interstellar medium with moderate efficiency (of order \( \sim 10\% \)), then provided a natural source of turbulent pressure capable of counteracting the large gas column densities \citep[e.g.][]{agertz2021,ejdet22}. 

In earlier papers \citep{bla24,bla25}, we explored gas-rich discs in unprecedented detail and over the full range of $f_{\rm gas}$ for the first time. 
We introduced our new high-resolution (parsec-scale) simulation framework, \textsc{Nexus}, designed to study the dynamical evolution of turbulent discs in early galaxies \citep{tep24}. 
The main properties of the ISM encoded into modern star formation theories are resolved, including the local Mach number ${\cal M}$, the turbulent flow parameter $b$, the cloud virial parameter $\alpha_{\rm vir}$ and so forth \citep{FederrathKlessen2012}.

Present-day Milky Way analogues, with low gas fractions (\( f_{\rm gas} \lesssim 10\% \), defined in Eq.~\ref{e:fdd}), offer only limited insight into the gas physics governing galaxies at earlier cosmic times. Gas-rich systems exhibit both intermittent and long-lived behaviour, including disc–halo interactions, the formation of bars and spiral arms, and the emergence of central bulges \citep{bla24}. Using this framework, we demonstrated for the first time that bar-like phenomena are expected across the full range of gas fractions, even in fully gas-dominated turbulent discs, and these are expected to be active sites of star formation, unlike bars today. Notably, we identified a distinctive ``radial shear flow'' whose characteristic signature develops prior to the formation of the central bulge. If the predicted kinematic signature is confirmed, this will argue strongly that {\it turbulent gas dynamics dominates the evolution of early discs}.

With every new ALMA or JWST observation, important insights have begun to emerge. Here we address just one of these $-$ the appearance of massive stellar bars ($M_{\rm bar}\gtrsim 10^{10}$ M$_\odot$) in gas-rich galaxies at $z\gtrsim 4$ \citep{sma23,boo26,wang26} when the Universe was in its first billion years. (To emphasize the extraordinary nature of these sources, their main properties are tabulated in Table~\ref{t:bars}.)

In conventional models, large massive bars take billions of years to grow, typically as the bar slows down due to its interaction with the dark matter halo through dynamical friction \citep{ath03}. Using a (gas-free) cold stellar disc with $10^8$ particles, \citet{fuj18a} demonstrated how a slowing bar's radius can grow at 0.6 kpc Gyr$^{-1}$ over the entire age of the Universe. At a fixed total disc (gas$+$stars) and halo mass, a higher fraction of dynamically inert gas progressively delays the emergence of the stellar bar \citep{ath13,bla23}.
Thus, a conventional bar-forming model cannot easily explain massive, high-redshift bars.\footnote{None of the $z\gtrsim 4$ disc systems show clear evidence for a major merger interaction to trigger such a bar, but possible companions can be identified.}

In contrast, we see a very different outcome in gas-rich discs supported by turbulence. The turbulent gas phase,  sustained by energetic feedback from supernovae and gravitational instabilities, is different in character from the inert, laminar gas flow in early studies \citep[e.g.][]{ath13}. This was apparent when comparing our inert gas \citep{bla23} and turbulent gas \citep{bla24} simulations {\it under identical disc-halo set-up conditions}, where the emergence of stellar bars is accelerated in turbulent gas discs.

Within the \textsc{Nexus} framework, we find that there are two key parameters that can explain the existence of early massive stellar bars:
(i) The disc mass fraction, $f_{\rm disc}$, which determines whether the disc baryons dominate the underlying dark matter halo,
\begin{eqnarray}
\label{e:fd}
    f_{\rm disc}= \left(\frac{V_{\rm c, disc}(R_{\rm e})}{V_{\rm c, tot}(R_{\rm e})}\right)_{R_{\rm e}=2.2 R_{\rm disc}}^2 \;.
\end{eqnarray}
Here, $V_c(R)$ is the circular velocity at a radius $R$, $R_{\rm disc}$ is the exponential disc scale length, and $R_{\rm e}=2.2 R_{\rm disc}$ is the traditional scale length adopted in studies of discs; (ii) The gas mass fraction within the same radial scale:
\begin{eqnarray}
\label{e:fdd}
    f_{\rm gas}=
    \left(\frac{M_{\rm disc, gas}}{M_{\rm disc}}\right)_{R_{\rm e}=2.2 R_{\rm disc}}
\end{eqnarray}
where $M_{\rm disc}$ is the total disc mass and $M_{\rm disc,gas}$ is the gas contribution.  

The first parameter has been explored extensively in recent years \citep{fuj18a,bla23,che25}. In seminal work, \citet{fuj18a} demonstrated that the bar onset time (defined when the normalized growing $m=2$ mode, $A_2/A_0$, exceeds 0.2 for the first time) has an inverse exponential dependence on $f_{\rm disc}$, with more dominant discs forming bars more rapidly (``Fujii relation''). This ingredient is suggested by mounting evidence for early star-forming discs dominating the local gravitational potential \citep{Price2021}.
But as the disc becomes more dominant, it forms a strong bar that buckles through vertical bending modes; these heat the inner disc, and ultimately destroy the bar \citep{com90,rah91}. 

The resolution to the buckling problem is to include a gas phase that tends to soften the impact of buckling \citep{ber98,deb06,ber07,ath13,lok20,lok25}. 
By raising $f_{\rm disc}$ from 50\% in our earlier work to 70\%, and varying $f_{\rm gas}$ at a fixed total disc mass and surface density over the full range of $f_{\rm gas}$ (see Table~\ref{t:mod}), we are able to produce massive stellar bars within a gigayear, as inferred from Table~\ref{t:bars}. 

In our new work, we therefore suggest that massive stellar bars arise at two distinct epochs: (i) at $z=0$ because they have grown slowly over billions of years, and (ii) at high redshift due to rapid bar growth in dominant gas-rich discs.
Section 2 presents the simulation and analysis methods. Section 3 introduces the new results and Section 4 describes the role of statistical physics in the emergence of bars.
Section 5 discusses the consequences of our new work for the latest high-redshift discoveries before summarizing our conclusions in Section 6.

\section{Simulations and analysis}

\subsection{Models and movies}

This work focuses on systems with halo masses of order $10^{11}$ ${\rm M}_\odot$, characteristic of the dynamical masses inferred for recent JWST discoveries at $z\sim1$--6 \citep{guo23,leconte24,costan23}. Such masses are also expected for Milky Way progenitors at $z\approx3$--5 \citep[Fig. 1 in][]{bla16}. At high redshift, the combination of high gas fractions and enhanced gas surface densities \citep{tac20} drives elevated star-formation rates. The resulting injection of energy and momentum, together with gravitational instabilities, couples to the gas and sustains a highly turbulent ISM \citep{agertz09b,agertz09,jim23}. The adopted star and galaxy formation physics is described in \citet{age13} and \citet{agertz2021}. Stellar feedback injects sufficient energy and momentum to maintain this turbulent support.

Our first step is to consider an isolated galactic ecosystem in dynamical equilibrium, both with and without smooth accretion from the ambient hot corona. 
All computations were carried out with the \ramses\ N-body/hydrodynamics code \citep[][]{tey02a} at benchmark ($N\sim 10^7$ elements) including star formation and metal production (see Table~\ref{t:mod}); the number of effective gas ``cells'' is a factor of 10 higher in both cases.
In defining a massive progenitor, we adopted a model with three key components: a live dark matter halo, a massive stellar/gaseous disc, and a hot coronal gas filling the live dark matter halo, which serves to supply the disc with a smooth flow of accreting gas after it cools.
For our massive progenitor, we adopted halo parameters of $R_{\rm vir}\approx 40$ kpc and $\log M_{\rm vir}/{\rm M}_\odot \approx 11$ consistent with the chosen epoch.

\begin{table*}
\setlength{\tabcolsep}{3pt}
\begin{tabular}{llcccc}
\hline
\: & Source                         & M1149-BSG-z5 & BRI 1335-0417 & SMG 850.1    & GN20 \\
\\
1.\: & Redshift                                   & 5.10      &   4.41         & 4.26       & 4.06 \\
2.\: & Lookback time (Gyr)                        & 12.6    &   12.4         & 12.3                & 12.2 \\
3.\: & Cosmic time (Gyr)                          & 1.2     &   1.4          & 1.5                 & 1.6 \\
4.\: & Stellar mass, $M_\star$ ($\log$ M$_\star/$M$_\odot$)  & 10.5    &   11.2?         & 11.6                  & 11.0 \\
5.\: & Gas mass, $M_{\rm gas}$ ($\log$ M$_\star/$M$_\odot$)      & 10.7?      &  11.7         & 11.0                  & 11.5 \\
6.\: & Dynamical mass, $M_{\rm dyn}$ ($\log$ M$_\star/$M$_\odot$)    & 11.2?   &  11.8     & 11.7                  & 11.7 \\
7.\: & Star formation rate ($\log$ M$_\odot$ yr$^{-1}$)    & 2.2    & 3.2         & 3.1                     & 3.3?  \\
8.\: & Gas fraction, $f_{\rm gas}$ (\%)            & 70$\pm$20$^\dag$   &  70$\pm$20         & 20$\pm$10         & 75$\pm$25 \\
9.\: & Disc fraction, $f_{\rm disc}$ (\%)            & high         &  75$\pm$20 & high           & 70$\pm$30 \\
10.\: & Effective radius, $R_{\rm e}$ (kpc)       & 2.6           &   3.4      & 3.8                 & 3.6 \\
11.\: & Bar radius, $R_{\rm bar}$ (kpc)              & 4.5         & 3.3        & 2.5                 & 2.8 \\
12.\: & Reconstructed lensed source & No & No & Yes & No \\
13.\: & Active galactic nucleus & Yes & Yes & No & Yes \\
\hline\\
\end{tabular}
\caption{Overview of the four earliest barred galaxies observed to date. The estimated values are from (Col. 2) \citet{wang26}, (Col. 3) \citet{tsu24} (Col. 4) \citet{sma23} and (Col. 5) \citet{boo26}. Notes: (Row 1) All redshifts were confirmed spectroscopically with timescales (Rows 2, 3) obtained from the 2024 Planck cosmology. The total stellar$+$gas masses (Rows 4, 5) are typically less than the rough dynamical mass estimates (Row 6); we caution that the stated masses have significant uncertainties of 0.15 dex or more. (Row 7) The star formation rates were determined from spectroscopy and photometry.
(Rows 8, 9) These were obtained using Eqs. 1 and 2, except for those indicated where $f_{\rm gas}$ was estimated from $^\dag$\citet{tac18} 
using Rows 4$-$6. The elevated gas fractions are supported by high dust content and extreme star formation rates ($> 100$ M$_\odot$ yr$^{-1}$) in Row 7. 
The effective radius in Row 10 has large uncertainties; Row 11 indicates the photometric bar radius.  All objects appear to be in overdense regions, one source is lensed (Row 12) and three show signs of an active nucleus (Row 13).}
\label{t:bars}
\end{table*}

\begin{table}
\begin{tabular}{cclccc}
\hline
$f_{\rm disc}$ & $f_{\rm gas}$ & Label & $M_\star$ & $M_{\rm gas}$ & $R_{\rm disc}$ \\
 &  &  & ($10^{11}$~\Msun) & ($10^{11}$~\Msun) & (kpc) \\
\hline
{\bf 0.5}  & 0.0 & fd50\_fg00\_nac & 0.112 &  0.0  &  1.8  \\
{\bf 0.5}  & 0.2 & fd50\_fg20\_nac & 0.088 &  0.022  &  1.8   \\
{\bf 0.5}  & 0.4 & fd50\_fg40\_nac & 0.067 &  0.044  &  1.8   \\
{\bf 0.5}  & 0.6 & fd50\_fg60\_nac & 0.044 &  0.067  &  1.8  \\
\hline
{\bf 0.5}  & 0.2 & fd50\_fg20\_ac & 0.088 &  0.022  &  1.8   \\
{\bf 0.5}  & 0.4 & fd50\_fg40\_ac & 0.067 &  0.044  &  1.8   \\
{\bf 0.5}  & 0.6 & fd50\_fg60\_ac & 0.044 &  0.067  &  1.8  \\
\hline
{\bf 0.7}  & 0.0 & fd70\_fg00\_nac & 0.555 &  0.0  &  3.0  \\
{\bf 0.7}  & 0.2 & fd70\_fg20\_nac & 0.444 &  0.111  &  3.0   \\
{\bf 0.7}  & 0.4 & fd70\_fg40\_nac & 0.333 &  0.222  &  3.0   \\
{\bf 0.7}  & 0.6 & fd70\_fg60\_nac & 0.222 &  0.333  &  3.0  \\
\hline
{\bf 0.7}  & 0.2 & fd70\_fg20\_ac & 0.444 &  0.111  &  3.0   \\
{\bf 0.7}  & 0.4 & fd70\_fg40\_ac & 0.333 &  0.222  &  3.0   \\
{\bf 0.7}  & 0.6 & fd70\_fg60\_ac & 0.222 &  0.333  &  3.0  \\
\hline\\
\end{tabular}
\caption{Overview of galaxy models. The DM host halo properties ($M_{\rm halo} = 10^{11}$~\Msun; $R_{\rm vir} = 37$ kpc; $r_{\rm s} = 9.2$ kpc) are identical across models and they describe a massive progenitor at $z \approx 3$. Table columns are as follows: (1) Disc-to-total mass ratio (Eq.~\ref{e:fd}); (2) Gas to total disc mass fraction at $t=0$ (Eq.~\ref{e:fdd}); (3) Model designation; (4) Disc stellar mass; (5) Disc gas mass; (6) Disc scalelength. Note that the total disc mass and surface density are constant across models.}
\label{t:mod}
\end{table}

In Table~\ref{t:mod}, simulations are distinguished by the different initial disc and gas fractions. At a fixed $f_{\rm disc} = 50\%$, we ran four models using $f_{\rm gas} = (0, 20, 40, 60)\%$. The filenames include the designation "{\tt nac}" because no accretion was allowed from the halo. A parallel set of models were also run with $f_{\rm gas} = (20, 40, 60)\%$ where halo cooling was permitted and these have the designation "{\tt ac}". 

For models with $f_{\rm gas}<100\%$, we included a pre-existing (`old') stellar disc such that $f_{\rm old}=1-f_{\rm gas}$. (For models with $f_{\rm gas} \approx 100\%$, we include a tiny pre-existing stellar population ($f_{\rm old}\approx 1\%$) as a useful tracer of the galaxy's gas-driven evolution.) The $f_{\rm disc} = 50\%$ disc parameters in Table~\ref{t:mod} are broadly consistent with arguably the best Milky Way progenitor analogue to date, the object CEERS-2112 at a photometric redshift of $z\approx 3.0$ \citep{costan23}.

So as to address the recent discoveries in Table~\ref{t:bars}, the entire set of 7 simulations was repeated with a fixed $f_{\rm disc} = 70\%$, thus there are 14 simulations listed in the table. (We also ran simulations at $f_{\rm gas} = (80, 100)\%$ but these form instantaneous bars that collapse in a gigayear to form central bulges \citep{bla24}.) Quick-look snapshot images taken from all simulations are presented in Figs.~\ref{f:lastbar} and \ref{f:lastbar2}.

Note that, for the "{\tt nac}" set, the total disc mass, scale radius and surface density are constant, and the initial scale length of the disc (both gas and stars) was roughly maintained across models. In all models, the gas fraction declines as more of the mass is locked up in stars \citep[see Fig. 3 in][]{bla24}. 
The total mass of stars produced in 2 Gyr from low to high gas fraction spans $2.6-6.2 \times 10^9$ M$_\odot$ for the accreting halo models, and $0.6-4.0 \times 10^9$ M$_\odot$ for the non-accreting models presented here.
The baryon mass was preserved in both the accreting and non-accreting halo models. In the former case, the total baryon mass of the disc increases with time. 

At our website,\footnote{ \href{http://www.physics.usyd.edu.au/turbo\_discs/}{http://www.physics.usyd.edu.au/turbo\_discs/}} we provide a series of movies (animations) that show the evolution of each of our model galaxies. There are two types of animations: 1) `on-the-fly' animations; and 2) `post-processed' animations.
The `on-the-fly' animations were created at simulation runtime; these have a high time resolution ($\delta t \approx 1$~Myr), and they show the evolution of the newly formed stars, of the gas density, and of the gas temperature on a face-on projection. The `post-processed' animations have been created from the simulation outputs; they have a lower time resolution ($\delta t \approx 10$~Myr), and they show the evolution of the surface density of all disc components (gas, pre-existing stars, and - if available - newly formed stars, gas) along three orthogonal projections.

\begin{figure}
\includegraphics[width=\columnwidth]{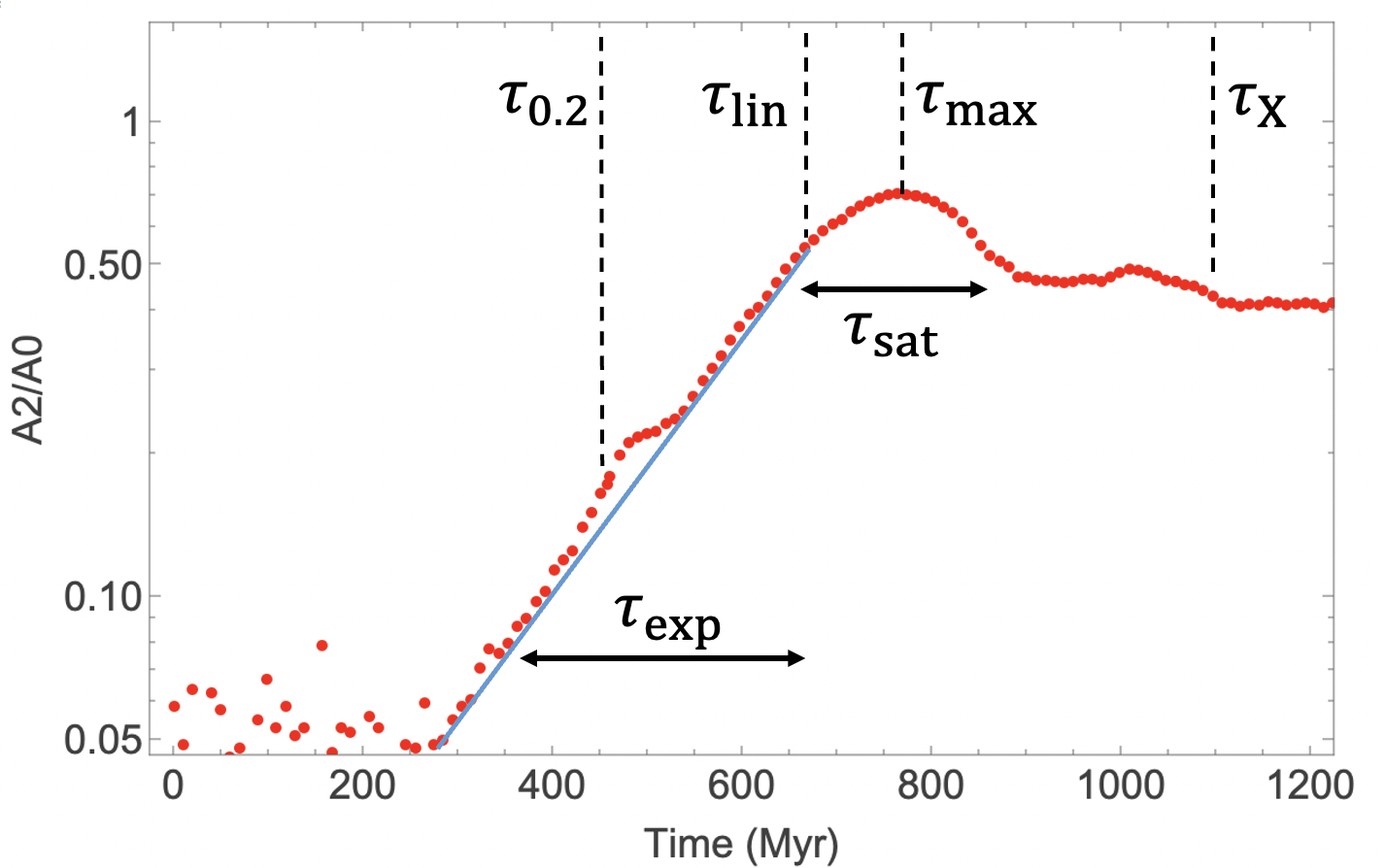}
\caption{Characteristic timescales in a bar's evolution traced by the normalised $m=2$ Fourier amplitude. The straight line emphasizes the exponential (linear) growth of density perturbations.}
    \label{f:times}
\end{figure}

\subsection{Dehnen's method}
We use Dehnen's method\footnote{Dehnen's method is a reference to his {\tt python} scripts; Fourier analysis of the simulated surface density is more common than analysis of the gravitational potential $\Phi$ because it is easier to relate to observations \citep{ath02}.}  \citep{dehnen23} across multiple simulated images to extract the bar's main parameters (bar amplitude, radius, etc.) at each timestep. The Dehnen method uses the fact that the continuity equation links the velocity field to the time derivative of the density pattern, so one can infer how fast the bar is rotating without needing to wait for it to rotate between snapshots. Thus, the measurements taken from each image are independent. 

But applying the Dehnen method to noisy simulations requires some caution. The method estimates the bar pattern speed $\Omega_{\rm bar}$ from the instantaneous time derivative of the $m=2$ Fourier phase, making it more sensitive to noise than methods based on measuring the phase evolution over a finite time interval. We have chosen to proceed with the Dehnen method to aid comparison with earlier stellar analyses, and this comparison serves to show that bars forming in gas-rich discs really are different. In a follow-up paper, Villa-Alatorre et al (2026) present a more robust approach to measuring bar parameters in gas-rich discs.

The disc's surface density $\Sigma(R)$ is decomposed into azimuthal Fourier modes \citep[e.g.][]{sel86}:
\begin{eqnarray}
A_m(R) = \frac{1}{M(R)} \sum_j m_j \, W_j(R)\, e^{-i m \phi_j},
\label{e:fourier}
\end{eqnarray}
where $R$ is the cylindrical radius, $\phi_j$ is the azimuthal angle of particle $j$, $m_j$ is the particle mass,
and $M(R)$ is the total mass in the radial bin. The normalized window function $W(R)$ (see \citet{dehnen23} for details) is chosen with care 
to ensure that a robust bar region ($R_{\rm inner}<R<R_{\rm outer}$) is selected and that unwanted noise is not introduced in the Fourier analysis.
The radial bar strength is defined as the normalized $m=2$ amplitude, i.e.
$A_2(R)/A_0(R)$; the phase of the $m=2$ mode is $\phi_2(R) = \frac{1}{2} \arg\left(A_2(R)\right)$. Within the bar region, $\phi_2(R)$ is approximately constant and defines the bar orientation. 

The end of the bar region and its radius is defined as where the normalized amplitude drops below the threshold $A'_{\rm 0.2} = A_2(R)/A_0(R)\approx 0.1$.
The bar radius $R_{\mathrm{bar}}$ is defined as the largest radius for which the $m=2$ phase remains coherent, i.e.
\begin{eqnarray}
\left| \phi_2(R) - \langle\phi_{\mathrm{bar}}\rangle \right| < \Delta \phi,
\end{eqnarray}
where $\langle\phi_{\mathrm{bar}}\rangle$
is the mean inner bar phase and $\Delta \phi < 10^\circ$. The small angle criterion tries to ensure that spiral-arm features do not contribute to the bar measurements.
Due to the continuity equation, $\Omega_{\rm bar}$ is evalluated directly from the single-snapshot positions and velocities of the particles in the bar region. Two contributions appear: the particles’ azimuthal streaming, weighted by $W(R)$, and a flux term proportional to $v_R\: \partial W/\partial R$, where $v_R$ is a star's radial velocity, which accounts for bias from particles entering and leaving the window selection as they move.

\begin{figure*}
\centering
\includegraphics[width=0.7\textwidth]{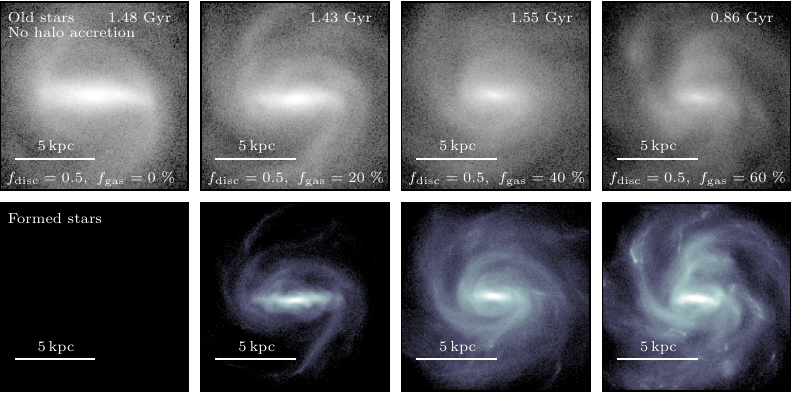}
\vspace{0.5cm}

\includegraphics[width=0.7\textwidth]{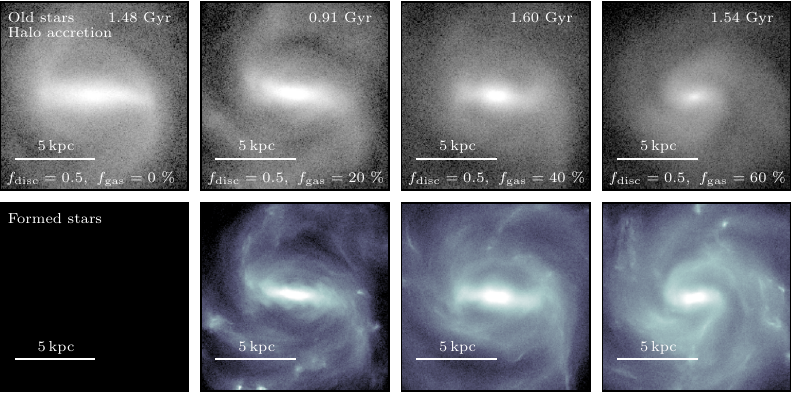}
    \caption{(Top) The four columns correspond to four $f_{\rm disc}=0.5$ simulations in order from the left, $f_{\rm gas}=0,20,40,60$\%. The snapshot times pick out a characteristic time when the bar is in the linear growth phase approaching peak amplitude. 
    There is {\it no} halo accretion active here; the box scale is 12$\times$12 kpc. The top row is the surface density of pre-existing stars; the bottom row is the surface density of created stars. 
    (Bottom) These are an identical set but {\it with} halo accretion activated. The intensity scalings are identical across the top and bottom panels.
    }
   \label{f:lastbar}
\end{figure*}
\begin{figure*}
    \centering
\includegraphics[width=0.7\textwidth]{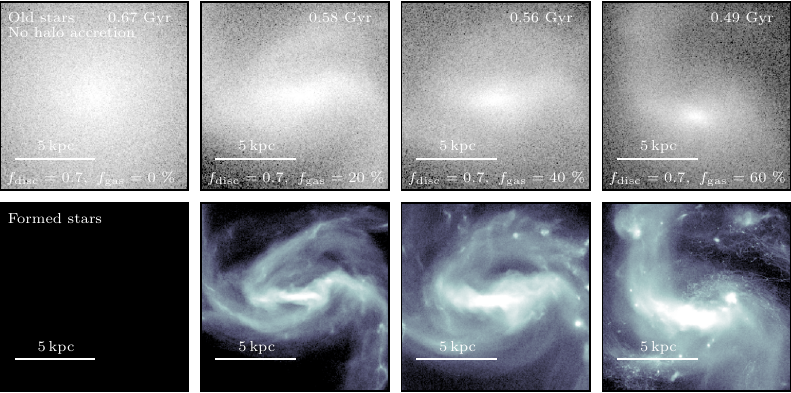}
\vspace{0.5cm}

\includegraphics[width=0.7\textwidth]{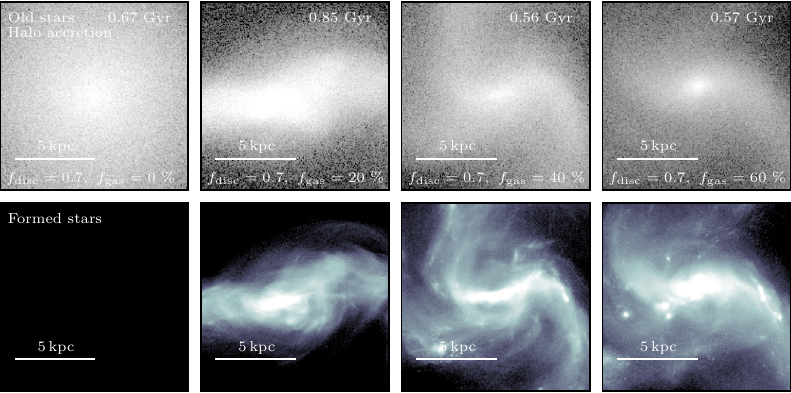}
     \caption{(Top) The four columns correspond to four $f_{\rm disc}=0.7$ simulations in order from the left, $f_{\rm gas}=0,20,40,60$\%. The snapshot times pick out a characteristic time when the bar is in the linear growth phase approaching peak amplitude. Note that for the $f_{\rm gas}=0\%$ case, there is no bar because of its disruption by vertical buckling (see text), in contrast to the $f_{\rm gas}>0\%$ cases that are stabilized by the gas.
    There is {\it no} halo accretion active here; the box scale is 12$\times$12 kpc.  The top row is the surface density of pre-existing stars; the bottom row is the surface density of created stars. 
    (Bottom) These are an identical set but {\it with} halo accretion activated. The intensity scalings are identical across the top and bottom panels. 
    }
   \label{f:lastbar2}
\end{figure*}
\begin{figure}
    \centering
    \begin{subfigure}{\columnwidth}
        \includegraphics[width=\columnwidth]{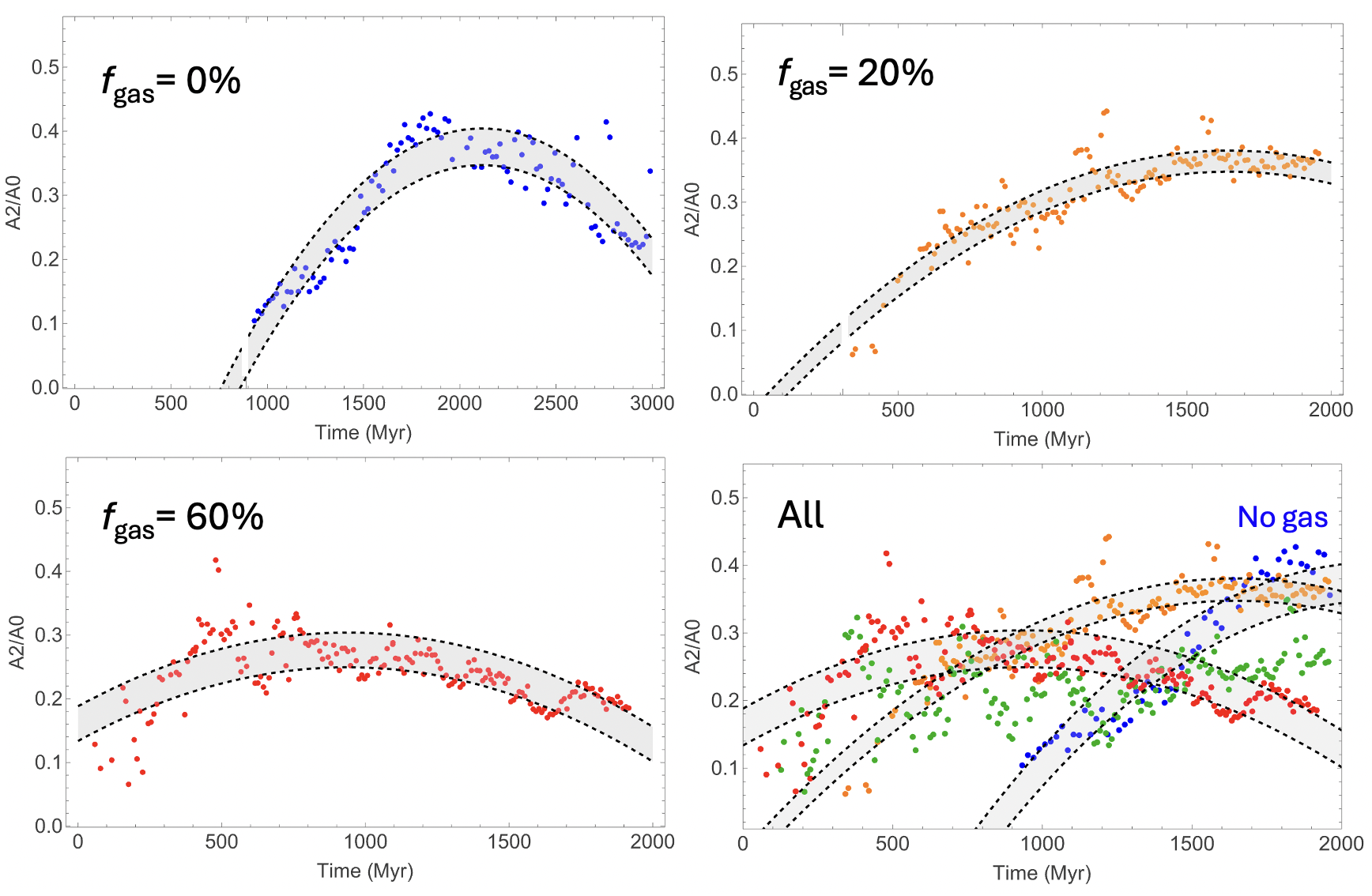}
        \caption{Bar strength, $A_2/A_0$}
        \label{f:f50A2}
    \end{subfigure}
    \vspace{0.5em}
    
    \begin{subfigure}{\columnwidth}
        \includegraphics[width=\columnwidth]{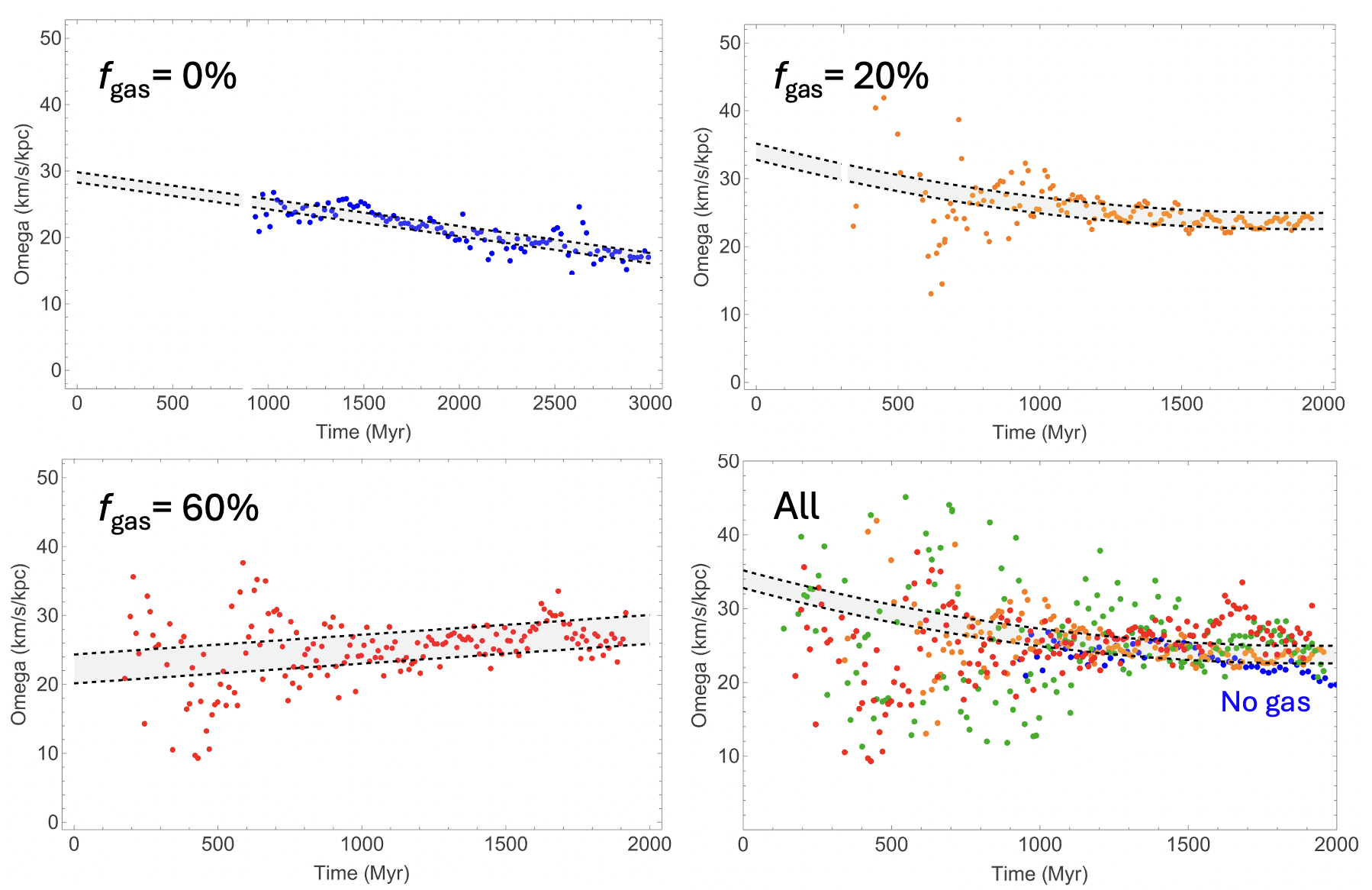}
        \caption{Bar pattern speed, $\Omega_{\rm bar}$}
        \label{f:f50Om}
    \end{subfigure}
    \vspace{0.5em}
    
    \begin{subfigure}{\columnwidth}
        \includegraphics[width=\columnwidth]{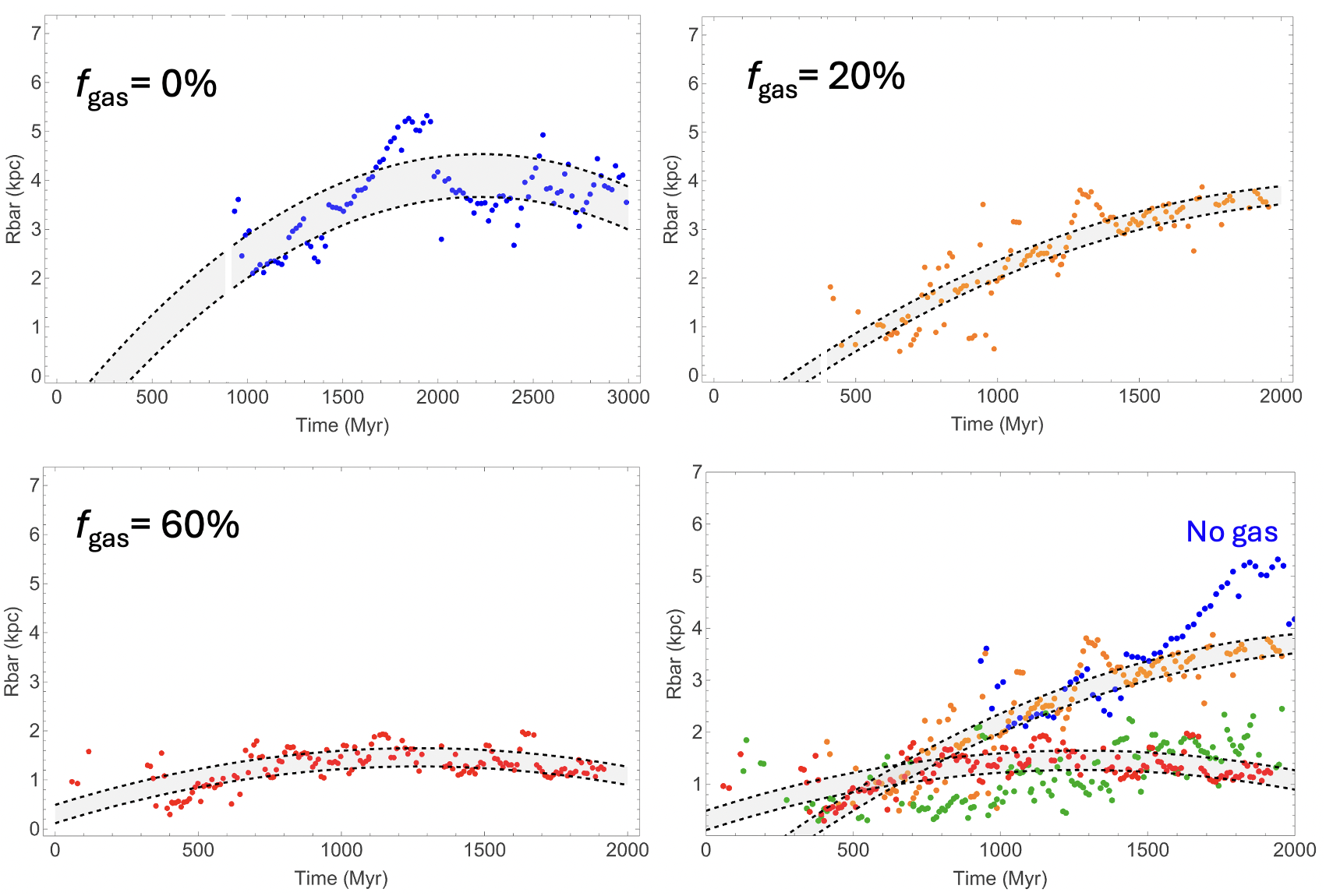}
        \caption{Bar radius, $R_{\rm bar}$}
        \label{f:f50Rbar}
    \end{subfigure}
    \caption{Dehnen's method applied to the $f_{\rm disc}=50\%$ models for varying $f_{\rm gas}$. To combat the noisy results, 50-percentile curves are shown after fitting the data with a low-order polynomial. In each panel, the bottom right figure is an overlay of all data. Note that the timespan shown is 2 Gyr in all figures, except $f_{\rm gas}=0\%$ where we present data for 3 Gyr due to the slower bar development.}
    \label{f:f50dehnen}
\end{figure}

\begin{figure}
    \centering
    \begin{subfigure}{\columnwidth}
        \includegraphics[width=\columnwidth]{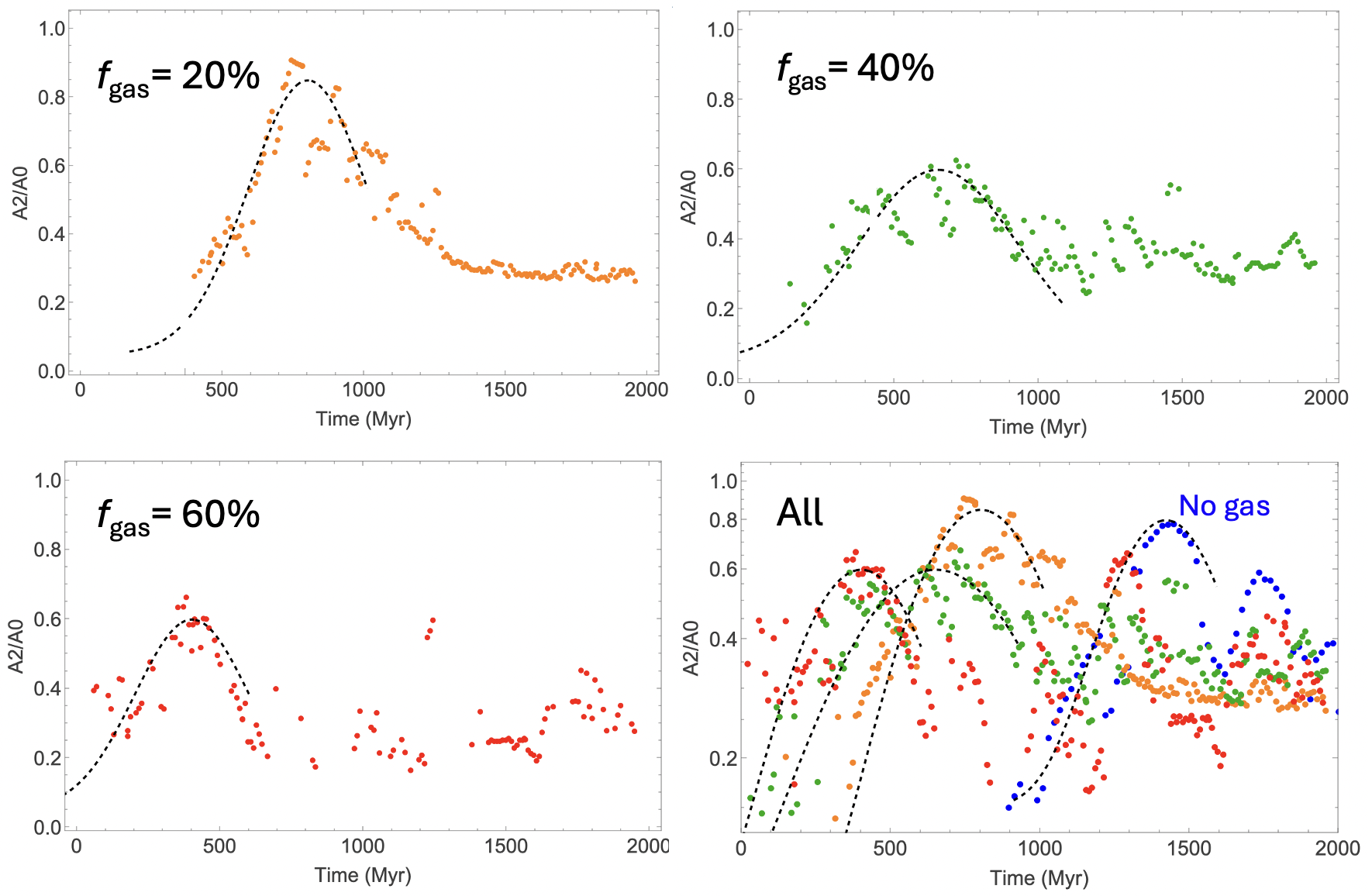}
        \caption{Bar strength, $A_2/A_0$}
        \label{f:f70A2}
    \end{subfigure}
    \vspace{0.5em}
    
    \begin{subfigure}{\columnwidth}
        \includegraphics[width=\columnwidth]{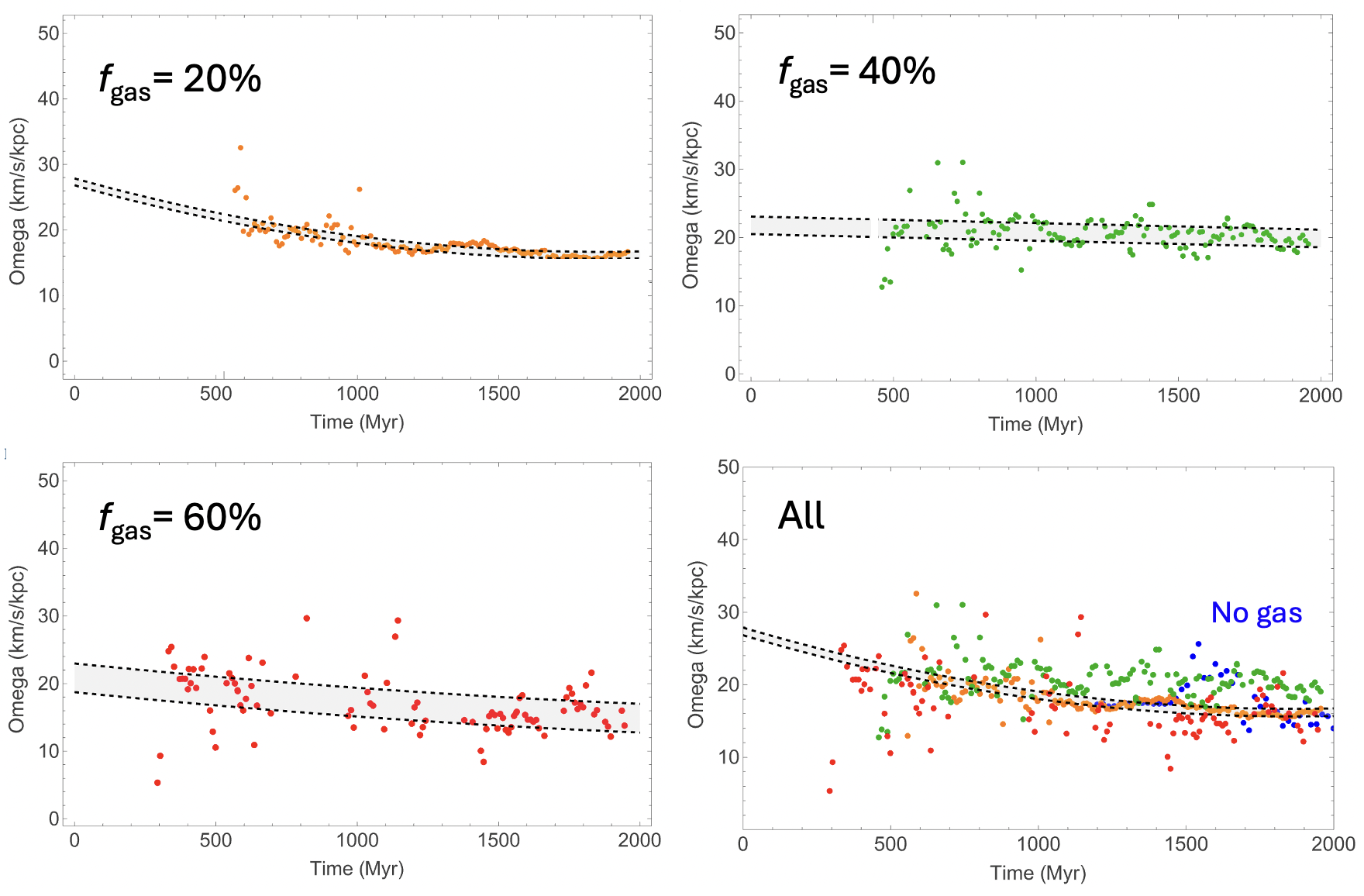}
        \caption{Bar pattern speed, $\Omega_{\rm bar}$}
        \label{f:f70Om}
    \end{subfigure}
    \vspace{0.5em}
    
    \begin{subfigure}{\columnwidth}
        \includegraphics[width=\columnwidth]{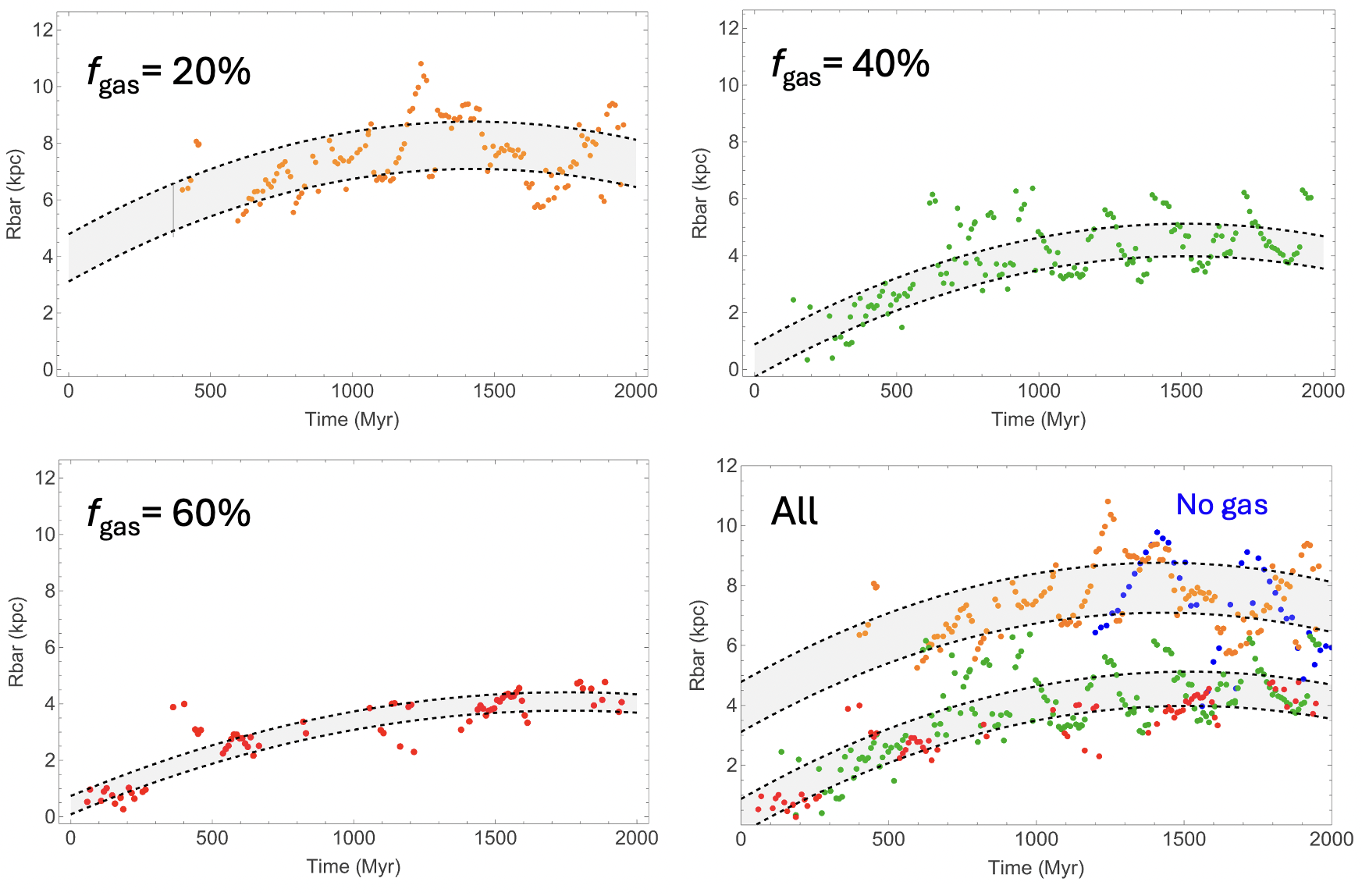}
        \caption{Bar radius, $R_{\rm bar}$}
        \label{f:f70Rbar}
    \end{subfigure}
    \caption{Dehnen's method applied to the $f_{\rm disc}=70\%$ models for varying $f_{\rm gas}$. To combat the noisy results, in (a), we gaussian-fit the rising curves to capture the peak amplitude. In (b) and (c), 50-percentile curves are shown after fitting the data with a low-order polynomial. In each panel, the bottom right figure is an overlay of all data.}
    \label{f:f70dehnen}
\end{figure}

\begin{figure}
\includegraphics[width=\columnwidth]{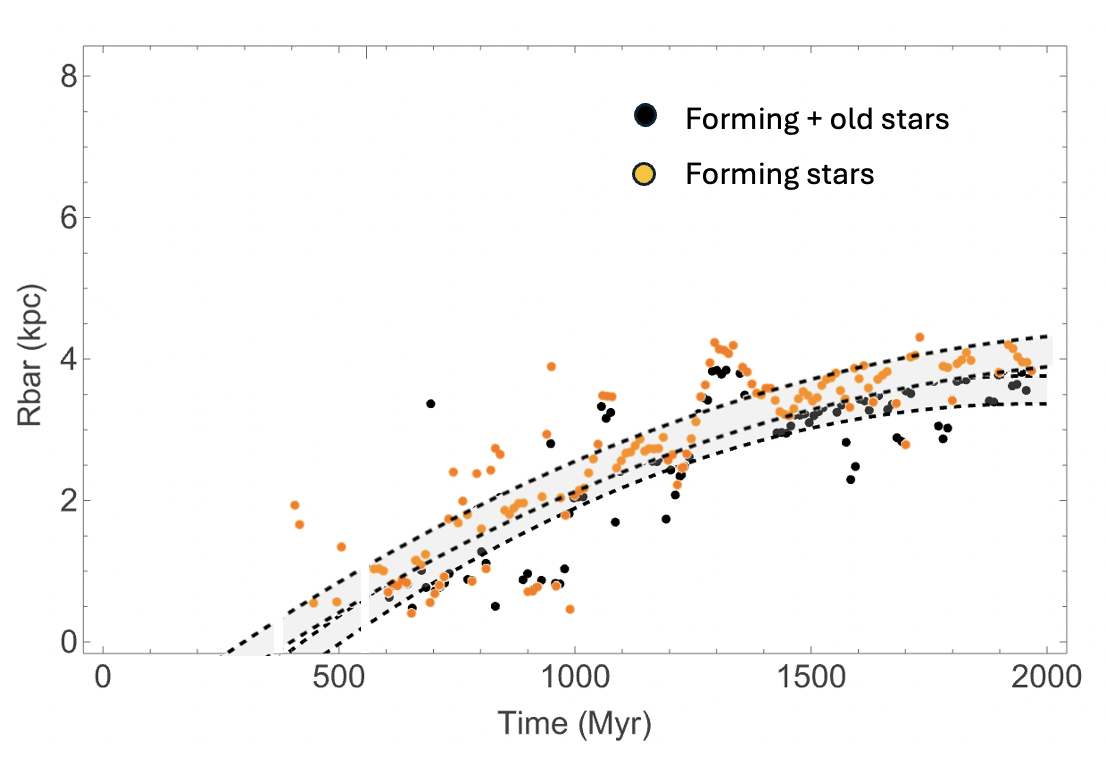}
\caption{The growth of the bar in the 
$(f_{\rm disc},f_{\rm gas})=(50\%,20\%)$ model as measured from the forming stars (orange points) and the pre-existing $+$ forming stars taken together (black points). The 50-percentile bands are essentially overlapping such that the bar's growth is {\it not} a consequence of any radial dependence of the star formation. We have chosen this model because it has the fastest rate of bar growth.
}
    \label{f:new_old}
\end{figure}
\begin{figure}
\includegraphics[width=\columnwidth]{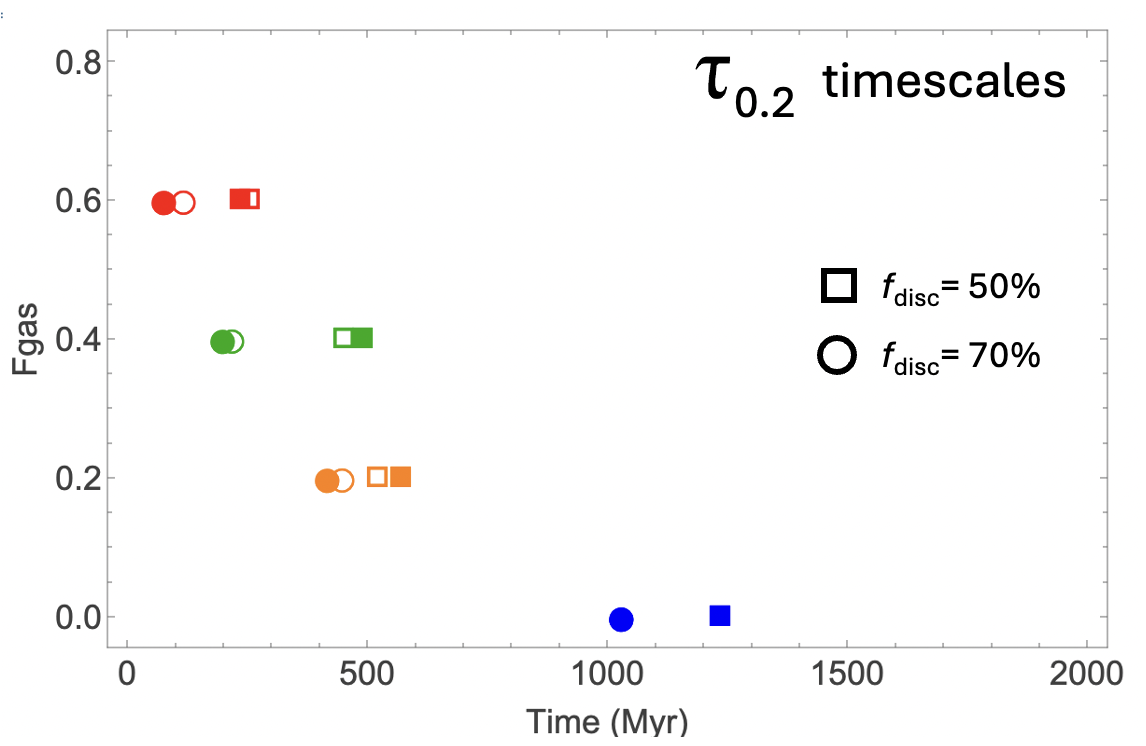}
\caption{The $\tau_{0.2}$ threshold timescale measured for all 14 models listed in Table~\ref{t:mod}. Filled symbols are for non-accreting halo models, empty symbols for the accreting models; $f_{\rm disc}$ is also indicated. The primary trend is that $\tau_{0.2}$ gets shorter as $f_{\rm gas}$ increases. There is also a weaker dependence on $f_{\rm disc}$ with $\tau_{0.2}$ decreasing as $f_{\rm disc}$ increases.
}
    \label{f:times2}
\end{figure}
\begin{figure}
\includegraphics[width=\columnwidth]{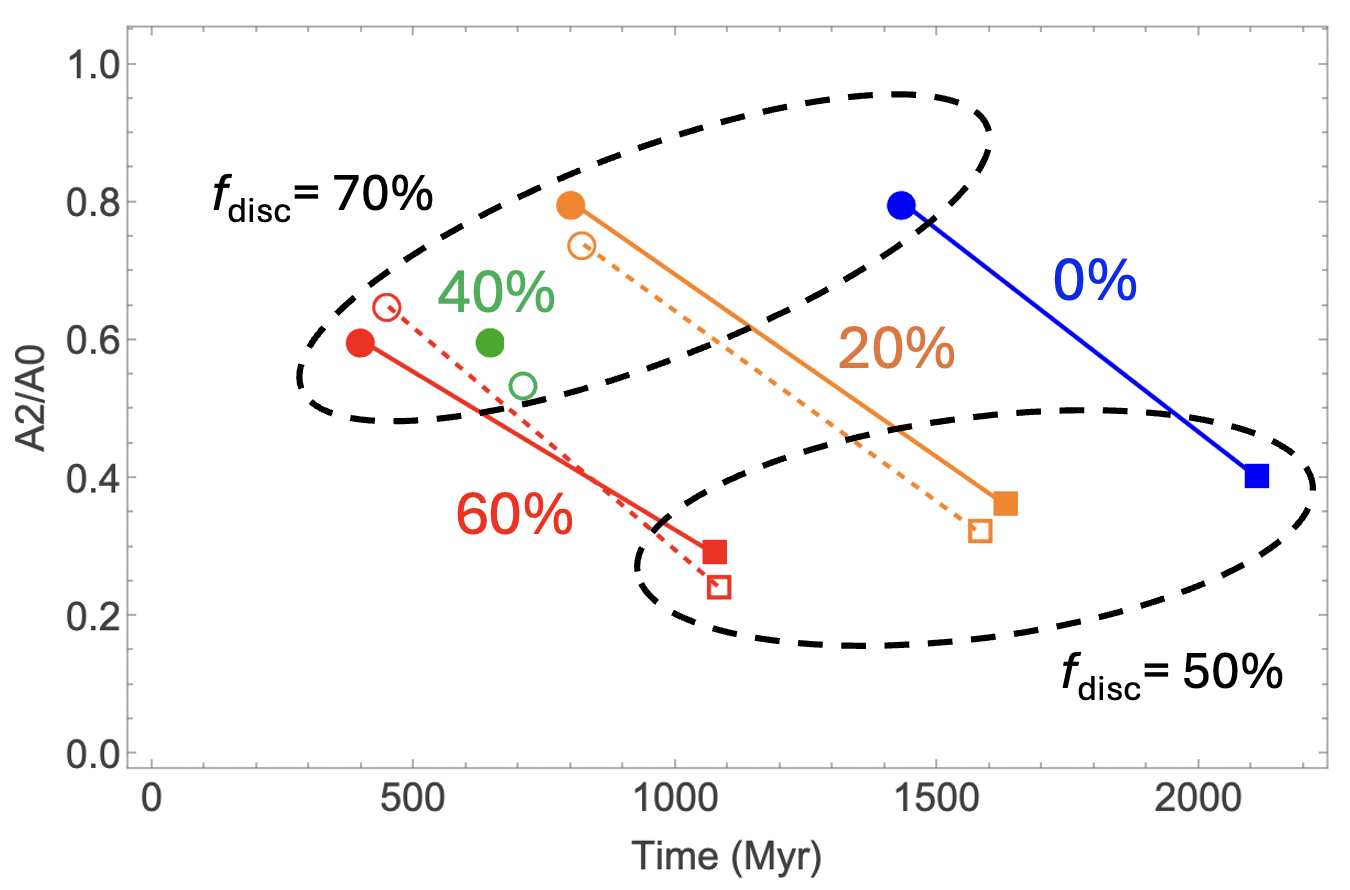}
\caption{For all models where measurable, the $\tau_{\rm max}$ timescales shift to shorter times when $f_{\rm disc}$ is increased from 50\% to 70\%, along with larger bar amplitudes, as illustrated. The symbols are defined in Fig.~\ref{f:times2}. The ($f_{\rm disc}=50\%$, $f_{\rm gas}=40\%$) models did not produce a well-defined peak amplitude. Fixed $f_{\rm gas}$ models are joined by solid (non-accreting) or dashed (accreting) lines.
}
    \label{f:times3}
\end{figure}


\subsection{Bar timescales}

The main focus of this work is to show how high gas fractions can expedite the formation (or onset) of a bar. But, in Fig.~\ref{f:times}, the concept of a ``bar formation'' time is somewhat arbitrary with different definitions used in past work. To illustrate the point, here we show the amplitude growth $A_2/A_0$ (log scale) plotted against the simulation time for a {\sc Nexus} simulation. It bears all the hallmarks of a disc simulation dominated by stars, with an exponential rise to a peak (saturation) amplitude followed by a weakening of the bar as it becomes vertically extended. The exponential fit to the initial rise is shown as a linear blue curve. 

There exists a variety of timescales involved in the exponential growth and saturation of a simulated bar.
Two measure durations: $\tau_{\rm exp}$ is the exponential or e-folding time \citep{bla23}; $\tau_{\rm max}$ is the saturation timescale measured at FWHM. The others are event times: 
$\tau_{\rm 0.2}$ is a widely used bar-formation time defined when the bar strength has reached a threshold value $A_2/A_0=0.2$ for the first time \citep{fuj18a,fujii19}; $\tau_{\rm lin}$ marks the end of the exponential linear-growth phase; $\tau_{\rm max}$ is the time of maximum amplitude; 
$\tau_{\rm X}$ is the time at which the disc starts to experience vertical bending modes, coinciding roughly with the first appearance of the vertical X-orbits that make up the boxy peanut bulge \citep{rah91,lok25}. 

In recent work, \citet{asano26} find that the time of maximum bar strength $\tau_{\rm max}$  is repeatable over many realizations, but $\tau_{\rm X}$ is not, being subject to chaotic orbit behaviour. These authors also confirm the validity of the exponential growth time, $\tau_{\rm exp}$.
An important event time (not shown) is the simulated zeropoint $\tau_0$ when the data are fitted with a function, e.g. $A_2/A_0 \propto e^{t-\tau_0}$, or when a stellar bar emerges within a cosmological simulation \citep{che25,frag25}.

\section{Results}

\subsection{Dehnen's method \& noisy data}

We now present the measured bar parameters for all simulations as a function of $f_{\rm gas}$. We focus on the bar strength ($A_2/A_0$), bar mass ($M_{\rm bar}$), bar radius ($R_{\rm bar}$), and bar pattern speed ($\Omega_{\rm bar}$). As foreshadowed, given the complex nature of gas simulations (Figs.~\ref{f:lastbar} and \ref{f:lastbar2}), this proved to be harder to achieve compared to traditional analyses of gas-free discs \citep[e.g.][]{fuj18a}.

When using the Dehnen method, the noise problem is acute for weak bars where the $m=2$ signal has low contrast and the phase becomes intrinsically noisy. Another problem arises from the intrinsic clumpiness of the evolving simulations. As $f_{\rm gas}$ increases, the gas and star-formation sites become increasingly clumpy as the disc evolves. This is particularly evident in the movies $-$ see Figs.~\ref{f:lastbar} and \ref{f:lastbar2}. 
Transient non-axisymmetric structure can introduce small-scale variations in the measured Fourier phase. The situation is further complicated when spiral structure or multiple non-axisymmetric components contribute significantly to the $m=2$ mode. In such cases, there may be no unique phase associated with the bar, and the instantaneous estimator can respond to changes in the relative contributions of different structures rather than to the physical rotation of the bar itself. Differentiation then amplifies these fluctuations, producing large excursions in the inferred parameters (e.g. pattern speed) even when the underlying bar is evolving smoothly. 

\textsc{Nexus} \citep{tep24} is the first code to allow for controlled, high-resolution, dynamically self-consistent calculations over the full range of gas fraction ($0\leq f_{\rm gas}\leq 100\%$). We stress that any related analysis that follows our new work will have to tackle the same sources of intrinsic noise.

\subsection{The emergence of a bar}

In Figs.~\ref{f:f50A2} and \ref{f:f70A2}, we compare and contrast the evolving bar amplitude $A_2/A_0$ for the $f_{\rm disc}=50\%$ and $f_{\rm disc}=70\%$ models respectively. We show only results for the ``non-accreting halo'' models because the ``accreting halo'' models look similar, albeit noisier. In most instances, the bar grows to a peak amplitude or strength before declining. The character of the rise and fall is different in both models, especially in terms of the bar's timescale to reach its maximum strength. For the $f_{\rm disc}=50\%$ models, the bar's growth is gradual and reaches lower peak amplitudes at the time of saturation. Note that all data are presented over 2 Gyr except for $f_{\rm gas}=0\%$ in Fig.~\ref{f:f50A2} presented over 3 Gyr to show the peak amplitude for the gas-free case. For the $f_{\rm disc}=70\%$ models, the onset of the bar is more rapid and higher bar amplitudes are reached.

To combat the noise, we fit low-order functions to the data to allow the different data sets to be compared. 
In Fig.~\ref{f:f50A2}, we use a parabolic fit in order to allow us to determine the 50-percentile band (shown in grey) about this curve. In the bottom right figure, we overlay three of these bands onto the data to guide the eye on how the peak amplitude shifts to earlier times as $f_{\rm gas}$ increases. Interestingly, the $f_{\rm gas}=40\%$ case has no well-defined peak amplitude.

In Fig.~\ref{f:f70A2}, the peaks are sharper, thus we fit a gaussian to the rising amplitude to capture the width, timescale and duration of the saturation peak. One again, in the bottom right figure, we overlay the four fits onto the data to guide the eye on how the peak amplitude shifts to earlier times as $f_{\rm gas}$ increases.

\subsection{The bar parameters}

In Fig.~\ref{f:f50Om} and Fig.~\ref{f:f70Om}, we compare the evolution of the bar's pattern speed for the $f_{\rm disc}=50\%$ and $f_{\rm disc}=70\%$ models respectively. What is striking is the similarity of the results across all models, within the measurement uncertainty. Once again,
we fit low-order functions to the data and define 50-percentile bands to allow the different data sets to be compared.

Most models show evidence for the emergent bar slowing down at later times, but the overall trend is weak over the 2 Gyr time frame compared to prior work \citep[cf.][]{ath03}.
Based on earlier work \citep[e.g.][]{davis26}, we had anticipated a stronger bar slow-down in the $f_{\rm disc}=50\%$ models compared to the $f_{\rm disc}=70\%$ models because the dark-matter halo is more dominant on the scale of the disc, but this is not seen. This may reflect the high gas fractions and a possible internal lag arising from hydrodynamics.

A useful method we used to verify the $\Omega_{\rm bar}$ determination was to transform the simulation to the ``measured'' bar rotation frame. The eye is sensitive to bars being fixed in time (well determined $\Omega_{\rm bar}$) or oscillating in this frame (unreliable $\Omega_{\rm bar}$). This approach enabled us to confirm that much of the observed noise in the data was intrinsic to the measurement, not to the bar's evolution.

In Figs.~\ref{f:f50Rbar} and \ref{f:f70Rbar}, much like the growth in bar amplitude, we see clear differences between the models in terms of the measured bar radius. Note that the vertical scales are different, in that the $f_{\rm disc}=70\%$ models produce bars that are almost twice as large for a fixed $f_{\rm gas}$, compared to $f_{\rm disc}=50\%$ models. The other clear trend is that the bar lengths are on average shorter as $f_{\rm gas}$ increases, as first discussed in \citet{bla24}. 

In Fig.~\ref{f:new_old}, we show how the
growth of the bar is {\it not} a consequence of any radial dependence of the star formation. We have chosen the $(f_{\rm disc},f_{\rm gas})=(50\%,20\%)$ model because it has the fastest bar growth. (The $(f_{\rm disc},f_{\rm gas})=(50\%,0\%)$ model asymptotes just beyond 2 Gyr.) We compare the bar properties determined from the forming stars (orange points) with the pre-existing $+$ forming stars (black points). The 50-percentile bands are essentially overlapping such that the bar's growth is the same.

\subsection{Early bars}

We have already seen how some parameters vary strongly (e.g. $R_{\rm bar}$) as a function of $f_{\rm disc}$ and/or $f_{\rm gas}$, while others are only weakly varying (e.g. $\Omega_{\rm bar}$). Here we are interested in identifying which of the models listed in Table~\ref{t:mod} are the most likely to produce early massive bars. This is the overarching goal of the present work. 

In Fig.~\ref{f:times2}, we present the popular $\tau_{0.2}$ threshold timescale measured for all 14 models listed in Table~\ref{t:mod}.
Filled symbols are for non-accreting halo models, empty symbols for the accreting models; $f_{\rm disc}$ is also indicated. 
We refrain from including the set of complementary figures to Figs.~\ref{f:f50dehnen} and \ref{f:f70dehnen} for the accreting models as they look similar in character to the non-accreting models.

All $f_{\rm disc}=70\%$ models are shifted to shorter timescales compared to $f_{\rm disc}=50\%$ models, consistent with the Fujii relation \citep{fuj18a,bla23,che25}. As is evident, there is no clear distinction between the accreting and non-accreting models at a fixed $f_{\rm disc}$ and $f_{\rm gas}$. The primary trend is that $\tau_{0.2}$ gets shorter as $f_{\rm gas}$ increases. There is also a weaker secondary dependence on $f_{\rm disc}$ with $\tau_{0.2}$ decreasing as $f_{\rm disc}$ increases.

While $\tau_{0.2}$ is widely used, in our view, it is more ``characteristic'' of rapidly growing ($f_{\rm disc}=70\%$) rather than slow-growing ($f_{\rm disc}=50\%$) bars. Thus, we include results for the peak amplitude time $\tau_{\rm max}$ as well. Interestingly, the trends are much the same.
In Fig.~\ref{f:times3}, the $\tau_{\rm max}$ timescales shift to shorter times when $f_{\rm disc}$ is increased from 50\% to 70\%, along with typically larger bar amplitudes, as illustrated. The ($f_{\rm disc}=50\%$, $f_{\rm gas}=40\%$) models do not produce a well-defined peak amplitude.

Note that, for $f_{\rm disc}=70\%$, both accreting and non-accreting models produce bars within 0.8 Gyr when $f_{\rm gas}\gtrsim 20\%$. This timescale drops to 0.4 Gyr when $f_{\rm gas}\gtrsim 60\%$, consistent with our goal to explain the $z\approx 5$ bar discovery \citep{wang26}.\footnote{In \citet{bla24}, it was shown that the $f_{\rm gas}= 80\%$ and $100\%$ cases fall into this category, but they collapse within 1 Gyr to form central bulges rather than sustaining a long-lived bar.} In a later section, we show that the estimated bar and disc masses are also consistent.

\section{The emergence of bars in fluctuating discs}

\subsection{Basic physics}

At the present time, there is no proper theory of bar formation, but there are several different approaches to describing the phenomenon.
Dynamically cold, rotating discs are low entropy, 
systems with excess (free) energy that can be extracted. A disc instability excites low-order modes that are fueled by gradients in the distribution function (DF) $f(J_i)$ where $J_i$ ($i: R, \phi, z$) are the dynamical actions.

In the linear regime, the instabilities grow at an exponential rate
\begin{eqnarray}
    A_2(t) = A_2(0) e^{\gamma_\star t}
    \label{e:amp}
\end{eqnarray}
where $A_2(0)$ is the amplitude of the triggered instability and the rate is given by 
\begin{eqnarray}
    \gamma_\star = \tau_{\rm exp}^{-1} \propto \left\vert\frac{\partial f}{\partial J_\mathrm{s}}\right\vert_\mathrm{res}
    \label{e:gamma}
\end{eqnarray}
where $\tau_{\rm exp}$ is the exponential (e-folding) growth time \citep{too74,bla23}.  
The proportionality on the
right-hand side is a generic feature of linear response theory. Its resonant character enters because, when
the full linear calculation is performed, the response integral is dominated by the neighbourhoods of a small
number of resonance locations, each resonance having an associated `slow' action $J_\mathrm{s}$.
The growth rate is then roughly proportional to the 
gradient of the DF along this slow action direction, evaluated at the resonance location. Energy and angular momentum transfer proceed as long as this gradient is preserved \citep{sel06,kod26}.

This picture also gives us an idea of when the linear phase will end, i.e. when the exponential growth will cease. This will be when the gradient of the DF near resonance is smoothed out, either by the redistribution of material caused by the instability itself, or by kicks from turbulent gas, or some other mechanism. At this point, the bar is said to have saturated at some maximum amplitude, $A^\prime_{\rm max}$. We stress that this is not an equilibrium configuration: further processes can develop (e.g. bar buckling) that cause the bar to evolve slowly, typically at lower amplitude (e.g. Fig.~\ref{f:times}).

The bar ultimately consists mostly of stars that are trapped onto $x_1$ orbits, whose apsides align along its long axis. It grows stronger over time by trapping more stars onto these bar-supporting orbits. 
Trapping is a complicated, inherently non-linear phenomenon, and so cannot be understood using the same linear theory that predicts the exponentially growing instability of Eqs.~\ref{e:amp} and \ref{e:gamma}.
However, what we can say robustly is that the bar will be better at nonlinearly trapping orbits (and hence at growing its own amplitude) as the galaxy's quadrupole moment increases.
Thus we expect that the larger $\gamma$, and the higher $A_2$ gets, the more efficient will be the trapping process, and hence the stronger will be the bar that ultimately forms \citep[q.v.][]{ham24}.

On the other hand, the above discussion does not take into account the role of (turbulent) gas. If the gas has a much lower turbulent velocity than $\sigma_\star$ then we might expect the self-gravitating response to be faster.
Also, the fact that there's a significant turbulent component means stars will feel significant stochastic fluctuations. 
The influence of a turbulent medium on bar formation is hard to predict. Naively one might
expect the stochasticity to act as described by \citet{hamilton2023galactic}: it preserves the gradient of the phase-space DF $\partial f/\partial J_\mathrm{s}$ near resonance, allowing angular momentum transfer to proceed for longer,
while simultaneously suppressing the trapping process. Turbulent gas would then assist the linear
physics and hinder the nonlinear physics.

The proposed physical mechanism is that bar formation occurs when the mode amplitude exceeds a critical value, enabling nonlinear trapping and subsequent growth. A higher gas fraction increases the fluctuation amplitude (parametrized here through an effective diffusion coefficient), which in turn 
enhances the probability of stochastic excursions beyond the critical amplitude barrier. Consequently, bar formation proceeds more rapidly in systems with larger gas content.

\subsection{What seeds the bar?}
\label{s:seed}

\begin{figure}
\centering
\includegraphics[width=0.8\columnwidth]{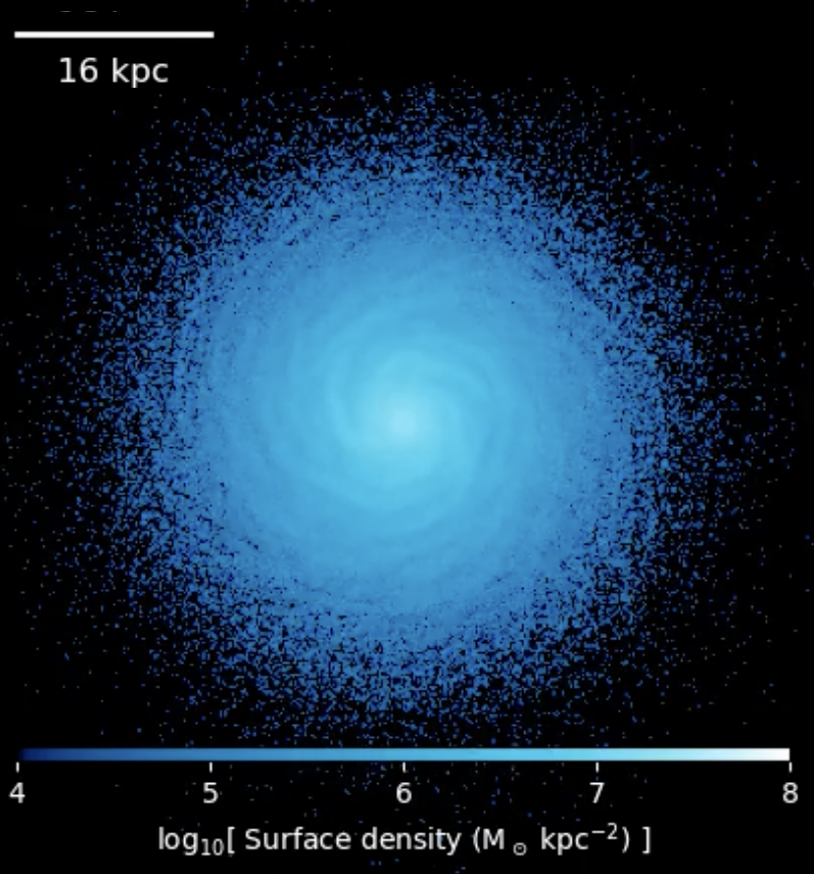}
\caption{The $(f_{\rm disc},f_{\rm gas})=(50\%,0\%)$ simulation was run for 4 Gyr. After 100 Myr, the disc generates flocculent or feather-like substructure with a normalized amplitude of a few percent (as shown here at $t=1$ Gyr) growing to $\approx 5\%$ at 4 Gyr. 
}
    \label{f:feather}
\end{figure}

The structure that seeds the bar in our
simulations sits well above the bare shot-noise level ($1/\sqrt{N} \lesssim 0.0003$), arising instead through swing amplification of the underlying
fluctuation field. Prior to bar formation, we measure a floor of
$\sigma_{A^\prime}\sim0.02-0.05$, and this structure is present even in the gas-free models (e.g. Fig.~\ref{f:feather}).
The illustrated $(f_{\rm disc},f_{\rm gas})=(50\%,0\%)$ model was run for 4 Gyr; over that timeframe, the simulated disc showed {\it no} indication of vertical heating \citep[][their Fig.~4]{bla21}. But after 100 Myr, the disc generated flocculent or feather-like substructure with an amplitude of a few percent growing to $\approx 5\%$ at 4 Gyr. Small leading/trailing density perturbations are swing-amplified by differential rotation. This azimuthal behaviour cannot be suppressed; even a disc that is locally cold and globally close to equilibrium is susceptible to self-excited, transient non-axisymmetric structure. In summary, {\it the bar is seeded by a swing-amplified background of smaller-scale fluctuations at the
few-percent level.}

Interestingly, our noise floor level (Figs.~\ref{f:times} and \ref{f:feather}) is the same as \citet[][their Fig. 7]{fuj18a} across all of their models. We identify this as arising from the azimuthal ``feathering'' effect seen clearly in their Figs. 2 and 3, i.e. high-order, multi-arm perturbations (or spurs) that are dynamically amplified \citep{fuj18a}. This was confirmed by an inspection of their simulations made possible by Dr. M. Fujii. The early rise from the Poisson floor at $t=0$ over a dynamical time to a similar higher background at later times is seen in \citet[][their Fig. 2]{Joshi24}. Regardless, the growth of structure is not related to the Poisson noise limit, contrary to earlier claims, although this may be the case in low-resolution ($N \lesssim 10^6$) simulations. In any event, here we consider the impact of physically-motivated fluctuating gas that is insusceptible to numerical Poisson noise as far as we can determine.

\subsection{Bar onset in a fluctuating disc}
\label{sec:onset_model}

\begin{figure}
    \includegraphics[width=\columnwidth]{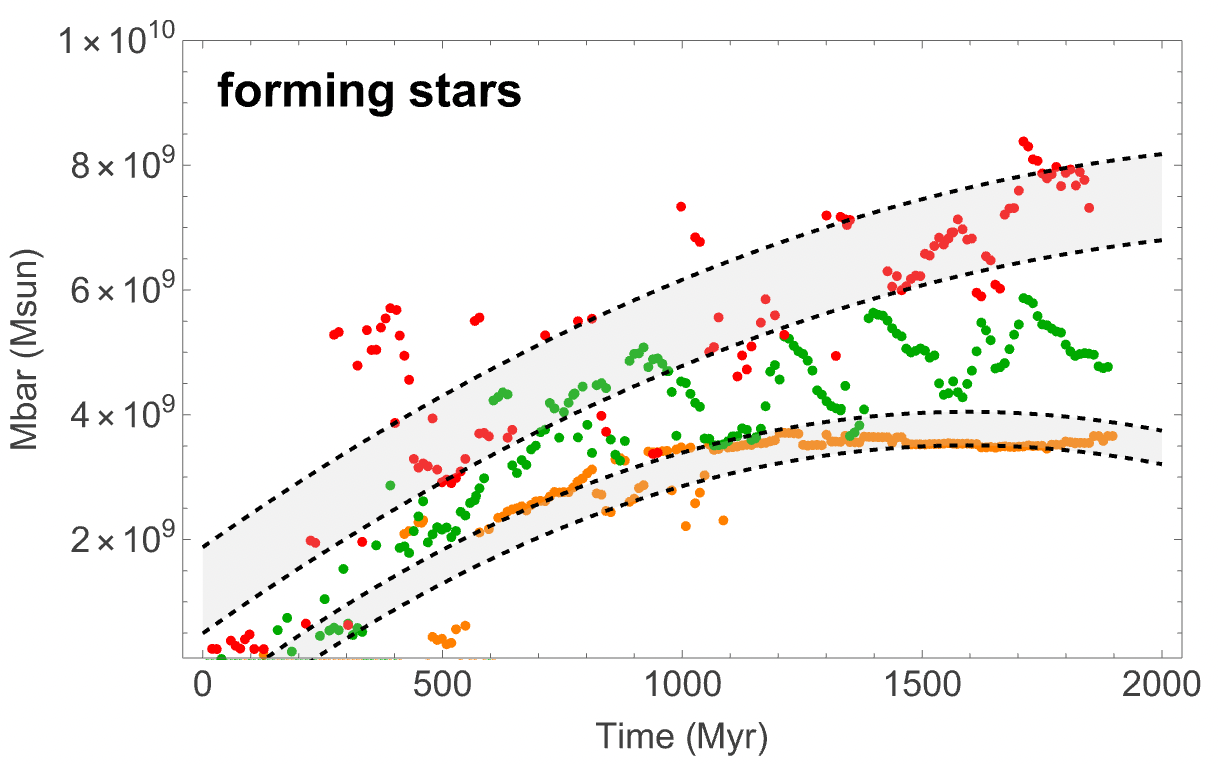}
    \includegraphics[width=\columnwidth]{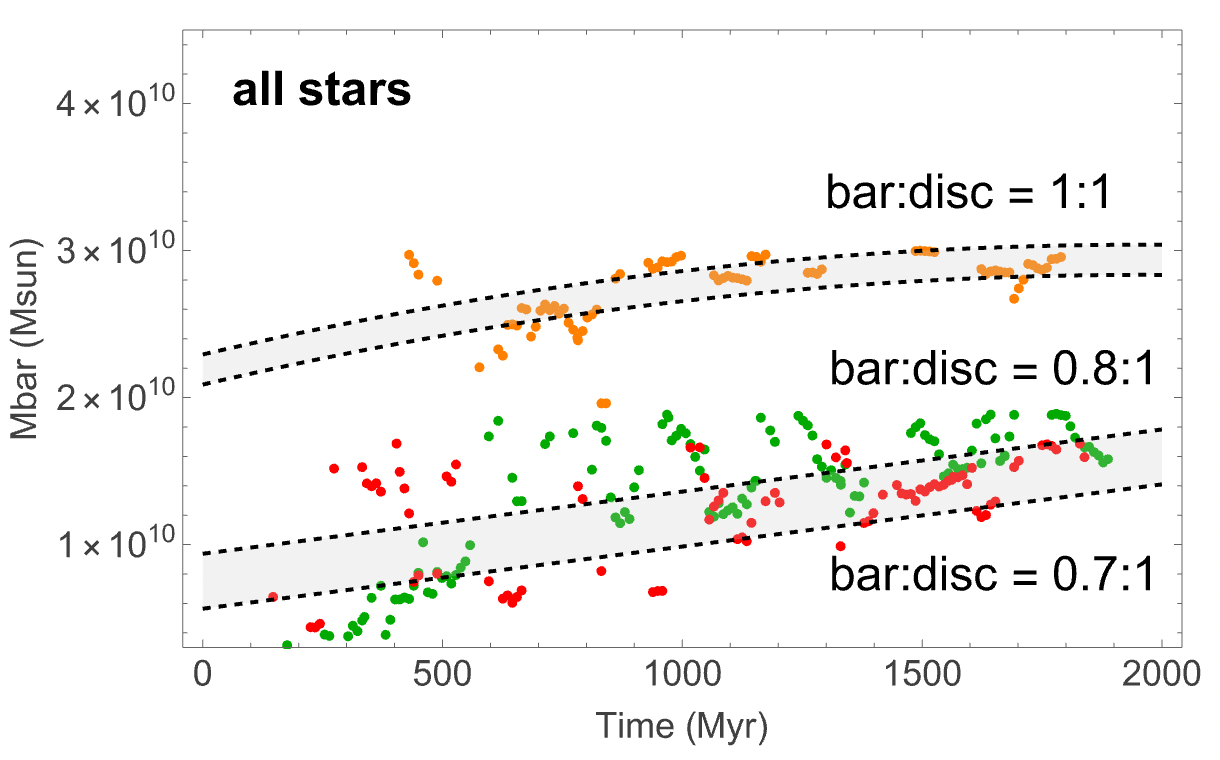}
    \caption{The evolution of the bar mass in forming stars (top) and pre-existing stars (bottom) with time. The colours represent different gas fractions, as described in Fig.~\ref{f:f50A2}. Once stabilised, the total bar mass is fairly constant and scales inversely with gas fraction. Note that the top and bottom figures have very different vertical scales. In the bottom figure, we indicate the approximate integrated bar-to-disc mass ratio out to $R=R_{\rm bar}$.}
    \label{f:Mbar}
\end{figure}
 
We now ask how a fluctuating gas component modifies the onset of the
bar instability. Our aim is deliberately modest: we do not attempt a
theory of bar formation, but rather a minimal description of how
stochastic forcing shifts the time at which a bar amplitude crosses a particular threshold.
 
In the \cite{zhang26} study of the \textsc{Nexus} simulations, their Fig.~3 shows the local density ($\rho$) fluctuation $\delta_{\rm gas}$ from the gas experienced by a star averaged over its orbit. An individual star does not respond to the instantaneous fluctuation, but to the ensemble average. A simple low-order fit at the disc's half-mass radius gives
\begin{eqnarray}
    \delta_{\rm gas} = \left\langle \frac{\vert\delta\rho\vert}{\rho} \right\rangle_{\rm orbit} \approx 0.8 f_{\rm gas} +0.04
    \label{e:dgas}
\end{eqnarray}
where, for consistency, we use the same form as \citet[][their Eq. 4]{zhang26}. It is convenient to work with the complex normalised amplitude
$a_2(t)\equiv A_2(t)/A_0$, with $A_2$ given by Eq.~\ref{e:fourier}, so that the
measured bar strength is $A_2' = |a_2|$. In the linear regime the coherent $m=2$
mode grows at the rate $\gamma_\star$ of Eq.~\ref{e:gamma} while the fluctuating
gas continuously injects power into the same azimuthal channel. The minimal
description of these two effects is
\begin{eqnarray}
\frac{{\rm d}a_2}{{\rm d}t} = \gamma_\star\,a_2 + \eta(t),
\qquad
\left\langle \eta(t)\,\eta^\ast(t')\right\rangle = D\,\delta(t-t') ,
\label{eq:langevin}
\end{eqnarray}
where $\eta$ is a complex forcing of zero mean representing the gas.
Averaging $|a_2|^2$ over realisations gives
\begin{eqnarray}
\frac{{\rm d}}{{\rm d}t}\left\langle A_2'^2 \right\rangle
= 2\gamma_\star \left\langle A_2'^2 \right\rangle + D ,
\label{eq:meansquare}
\end{eqnarray}
the first term describing growth of the mode and the second the
stochastic pumping by the gas. 
We assume throughout that $\gamma_\star$ is unchanged by the stochastic
forcing.
 
The single number $D$, with dimensions of (amplitude)$^2$ per unit
time, is the rate at which the fluctuations pump the mean-square $m=2$
amplitude. Rather than attempt to compute it from first principles, we
treat it as the one parameter characterising the gas. If the forcing
of the mode amplitude has root-mean-square value $\eta_{\rm rms}$ and
correlation time $\tau_{\rm corr}$, then $D\sim\eta_{\rm
rms}^2\,\tau_{\rm corr}$ up to a factor of order unity; treating the
forcing as white noise requires $\tau_{\rm corr}\gamma_\star \ll 1$.
Equation~(\ref{eq:meansquare}) integrates immediately to
\begin{eqnarray}
\left\langle A_2'^2 \right\rangle
= A_{\rm eff}^2\, e^{2\gamma_\star t} - \frac{D}{2\gamma_\star},
\qquad
A_{\rm eff}^2 \equiv A_2'^2(0) + \frac{D}{2\gamma_\star}.
\label{eq:solution}
\end{eqnarray}
The physical content of Eq.~(\ref{eq:solution}) is simple: the
fluctuations renormalise the
seed from which the mode grows. When
$D/2\gamma_\star \gg A_2'^2(0)$ the instability effectively starts
from a stochasticity-generated floor rather than from whatever coherent
perturbation was present initially.

In fact we can do better than solve for the mean-square amplitude alone.
Writing
$a_2(t) = b(t)\,e^{\gamma_\star t}$, the solution to the full Eq.~(\ref{eq:langevin}) is
\begin{eqnarray}
b(t) = a_2(0) + \int_0^t e^{-\gamma_\star t^\prime}\,\eta(t^\prime)\,{\rm d}t^\prime ,
\label{eq:seed}
\end{eqnarray}
We see that each kick from the gas enters $b$ with a weight $e^{-\gamma_\star t^\prime}$, so early
kicks count for far more than late ones --- a kick delivered early has the whole
subsequent growth phase in which to be amplified, one delivered near the end has
almost none. The integral therefore stops accumulating after a few e-folding
times and $b$ settles to a constant, $b_\infty$.
Given the statistics of $\eta$, and provided the forcing dominates any coherent
perturbation present initially ($|a_2(0)|^2 \ll D/2\gamma_\star$), one can show
that $|b_\infty|$ follows a Rayleigh distribution with
$\langle|b_\infty|^2\rangle = A_{\rm eff}^2$.
Once the initial phase is
over, every realisation simply grows exponentially from its own fixed
seed; but that seed is a random number, different in each realisation.

By the time $\gamma_\star t \gtrsim 1$, we have $b\simeq b_\infty = $ const.
Then the time at which a given realisation first reaches a threshold amplitude
$A_{\rm c}$ is
\begin{eqnarray}
\tau_{\rm onset} \simeq \frac{1}{\gamma_\star}\,
\ln\!\left(\frac{A_{\rm c}}{|b_\infty|}\right) ,
\label{eq:tauonset}
\end{eqnarray}
which is a random variable due to $b_\infty$. Averaging over the
Rayleigh distribution of $|b_\infty|$ gives
\begin{eqnarray}
\left\langle \tau_{\rm onset}\right\rangle = \tau_{\rm exp}
\left[\ln\!\left(\frac{A_{\rm c}}{A_{\rm eff}}\right)+\frac{\gamma_{\rm E}}{2}\right],
\qquad
\sigma_\tau = \frac{\pi}{2\sqrt{6}}\,\tau_{\rm exp} \simeq 0.64\,\tau_{\rm exp},
\label{eq:taustats}
\end{eqnarray}
where $\gamma_{\rm E}\simeq0.577$ is the Euler--Mascheroni constant and
$\tau_{\rm exp}=\gamma_\star^{-1}$.
Interestingly, the scatter $\sigma_\tau$ does not depend on the strength of
the stochastic forcing. The reason is that $D$ sets the \emph{scale} of the seed
distribution but not its shape: the fractional spread of a Rayleigh distribution
is a pure number, whatever its scale. Since the onset time depends on the seed
only logarithmically, a fixed fractional spread in $|b_\infty|$ produces a fixed
absolute spread in time. A larger $D$ therefore shifts the mean onset time
earlier and leaves the scatter untouched.

 
Adopting the
conventional $A_{\rm c} = A_{0.2}' = 0.2$, together with the pre-bar plateau
$\sigma_{A'}\approx0.02-0.05$ measured in \S\ref{s:seed},
Eq.~(\ref{eq:taustats}) gives
 \begin{eqnarray}
\tau_{0.2} \;\approx\; \left(1.7 - 2.6\right)\,\tau_{\rm exp}
\;\pm\; 0.64\,\tau_{\rm exp}
\label{eq:numerical}
\end{eqnarray}
i.e., a bar should appear after about two e-folding times of the
underlying instability, whatever the gas content, with a
realisation-to-realisation scatter of roughly two-thirds of an e-folding time.
 
The gas fraction enters Eq.~(\ref{eq:taustats}) only through
$A_{\rm eff}$ and, because the dependence is logarithmic, this is a
weak lever. Sweeping the floor across its entire measured range, from
$0.02$ to $0.05$, changes $\gamma_\star\tau_{0.2}$ by a factor of only $\approx1.5$. The measured onset times, by comparison,
fall from $\approx0.8$~Gyr at $f_{\rm gas}\gtrsim20\%$ to
$\approx0.4$~Gyr at $f_{\rm gas}\gtrsim60\%$ (\S3.4), a factor of two.
The conclusion is that stochastic seeding can account for most of the
observed shortening, but not quite all of it.
 
As a simple consistency check, it is worth seeing that the strength of stochastic fluctuations we have 
assumed is physically plausible. A fractional $m=2$ overdensity $\epsilon$ in
the inner disc perturbs the mode amplitude by $\sim\epsilon$ per
dynamical time, so $\eta_{\rm rms}\sim\epsilon\,\Omega$
and $A_{\rm eff}\sim\epsilon\,\Omega\,(\tau_{\rm corr}/
2\gamma_\star)^{1/2}$. Crucially, $\epsilon$ is \emph{not} the local
density contrast $\delta_{\rm gas}$ of Eq.~\ref{e:dgas}: only the
large-scale, quadrupolar part of the fluctuation spectrum couples to
the mode, so we write $\epsilon = \zeta\,\delta_{\rm gas}$ with
$\zeta < 1$ an unknown projection efficiency. Taking
$\delta_{\rm gas}\approx 0.5$, $\Omega\approx 0.07\,{\rm Myr^{-1}}$,
$\tau_{\rm corr}\approx 10\,$Myr and
$\gamma_\star^{-1}\approx 200\,$Myr, reproducing the observed
$A_{\rm eff}\approx0.03$ requires $\zeta$ of a few per cent. In other
words the measured floor is consistent with a small but not
implausible fraction of the turbulent power projecting onto the
coherent quadrupole.
 
We should stress what this model does not do. Most importantly, it assumes
$\gamma_\star$ to be independent of $f_{\rm gas}$, whereas 
a dynamically cold gas component will change the disc's linear response.
Beyond that, it says nothing about bar length or mass; it treats $A_{\rm c}$ as fixed, whereas 
diffusion plausibly raises the amplitude
required for trapping, so that $A_{\rm c}$ should itself increase with
gas fraction: it assumes white noise,
which requires $\tau_{\rm corr}\gamma_\star\ll1$, and neither
$\tau_{\rm corr}$ nor its dependence on $f_{\rm gas}$ has been
measured here. And the projection efficiency $\zeta$ relating the
local density contrast to the coherent quadrupolar forcing is
unknown; establishing it would require measuring the $m=2$ component of the gas potential in the inner disc directly.

\section{Discussion}

\subsection{Extreme discs}

We now focus our attention on the $f_{\rm disc}=70\%$ results as these are specifically relevant to the \citet{wang26} discovery, especially when we compare Fig.~\ref{f:times3} to Table~\ref{t:bars}. 
The new models constitute unchartered territory in the sense we have taken a new direction in computational research $-$ a study of physical processes in extreme dominant disc, gas-rich galaxies.

In Figs.~\ref{f:lastbar} and \ref{f:lastbar2}, we note the inverse dependence with $f_{\rm gas}$ in that more gas-rich bars are less massive on average. This is borne out in Fig.~\ref{f:f70Rbar} where the bar radius scales inversely with gas fraction, the same dependence observed for the $f_{\rm disc}=50\%$ \textsc{Nexus} models (Fig.~\ref{f:f50Rbar}). Thus, as concluded in our initial findings \citep{bla24}, the new quantitative analysis shows clearly that {\it more gas-rich galaxy discs form smaller stellar bars.}

\subsection{Bar masses}

In Fig.~\ref{f:Mbar} (bottom), the total bar masses (summed over pre-existing and forming stars) are $4\times 10^{10}$ M$_\odot$ ($f_{\rm gas}=0\%$; not shown),
$3\times 10^{10}$ M$_\odot$ ($f_{\rm gas}=20\%$; orange),
$1.8\times 10^{10}$ M$_\odot$ ($f_{\rm gas}=40\%$; green),
$1.5\times 10^{10}$ M$_\odot$ ($f_{\rm gas}=60\%$; red). 

In Fig.~\ref{f:Mbar} (top), we see that the early bars have ongoing star formation over the simulated timeframe. The star formation rates are higher as $f_{\rm gas}$ increases. By comparing both figures, it is evident that $f_{\rm gas}=60\%$ bars are much bluer than $f_{\rm gas}=20\%$ bars. But both are expected to show signs of active star formation, quite unlike their present day, red and dead counterparts.

Fig.~\ref{f:Mbar} (bottom) quotes the integrated bar-to-disc mass ratio out to $R=R_{\rm bar}$. In Table~\ref{t:bars}, all quoted masses are photometric, i.e. inferred from modelling the spectral energy distribution observed by JWST. The quoted masses are the total (baryonic) disc masses and thus they do not separate the bar from the underlying disc.

For ease of comparison with observations, the {\sc Nexus} simulated disc masses (pre-existing stars $+$ forming stars $+$ gas) need to be corrected for the gas fraction and the disc:bar mass ratio returned by the Dehnen code. The integrated mass (disc$+$bar, gas$+$stars) out to $R=R_{\rm bar}$ is roughly 
\begin{eqnarray}
    M_{\rm tot} = (1-f_{\rm gas})^{-1}(1+\frac{\rm\small disc}{\rm\small bar})M_{\rm bar} \;\:\:\:\:\: (f_{\rm gas} < 1, M_{\rm bar}>0).
\end{eqnarray}
Thus, the simulated total disc masses (10$-$20\% uncertainty) to compare with observations are:
$7.5\times 10^{10}$ M$_\odot$ ($f_{\rm gas}=20\%$),
$6.8\times 10^{10}$ M$_\odot$ ($f_{\rm gas}=40\%$),
$9.1\times 10^{10}$ M$_\odot$ ($f_{\rm gas}=60\%$), or logged values of 10.9, 10.8 and 11.0 ($\log$ M$_\odot$) respectively. 

In a few instances, where $z\gtrsim 4$ discs have ALMA observations, the stated masses are inferred dynamically (Table~\ref{t:bars}), such that
\begin{eqnarray}
    M_{\rm dyn} = (1-f_{\rm gas})^{-1}f_{\rm disc}^{-1}(1+\frac{\rm\tiny disc}{\rm\tiny bar})M_{\rm bar} \;\:\:\:\:\: (f_{\rm gas} < 1, M_{\rm bar}>0)
\end{eqnarray}
which includes a correction for the dark matter contribution inside 
$R=R_{\rm bar}$. (Note that if $M_{\rm bar}>0$, then $f_{\rm disc} >0$ and the disc:bar ratio is finite.) The simulated dynamical masses (20\% uncertainty) to compare with observations are:
$1.1\times 10^{11}$ M$_\odot$ ($f_{\rm gas}=20\%$),
$9.6\times 10^{10}$ M$_\odot$ ($f_{\rm gas}=40\%$),
$1.3\times 10^{11}$ M$_\odot$ ($f_{\rm gas}=60\%$), or logged values of 11.0, 11.0 and 11.1 ($\log$ M$_\odot$) respectively.

\begin{figure*}
\centering
    \includegraphics[width=1.4\columnwidth]{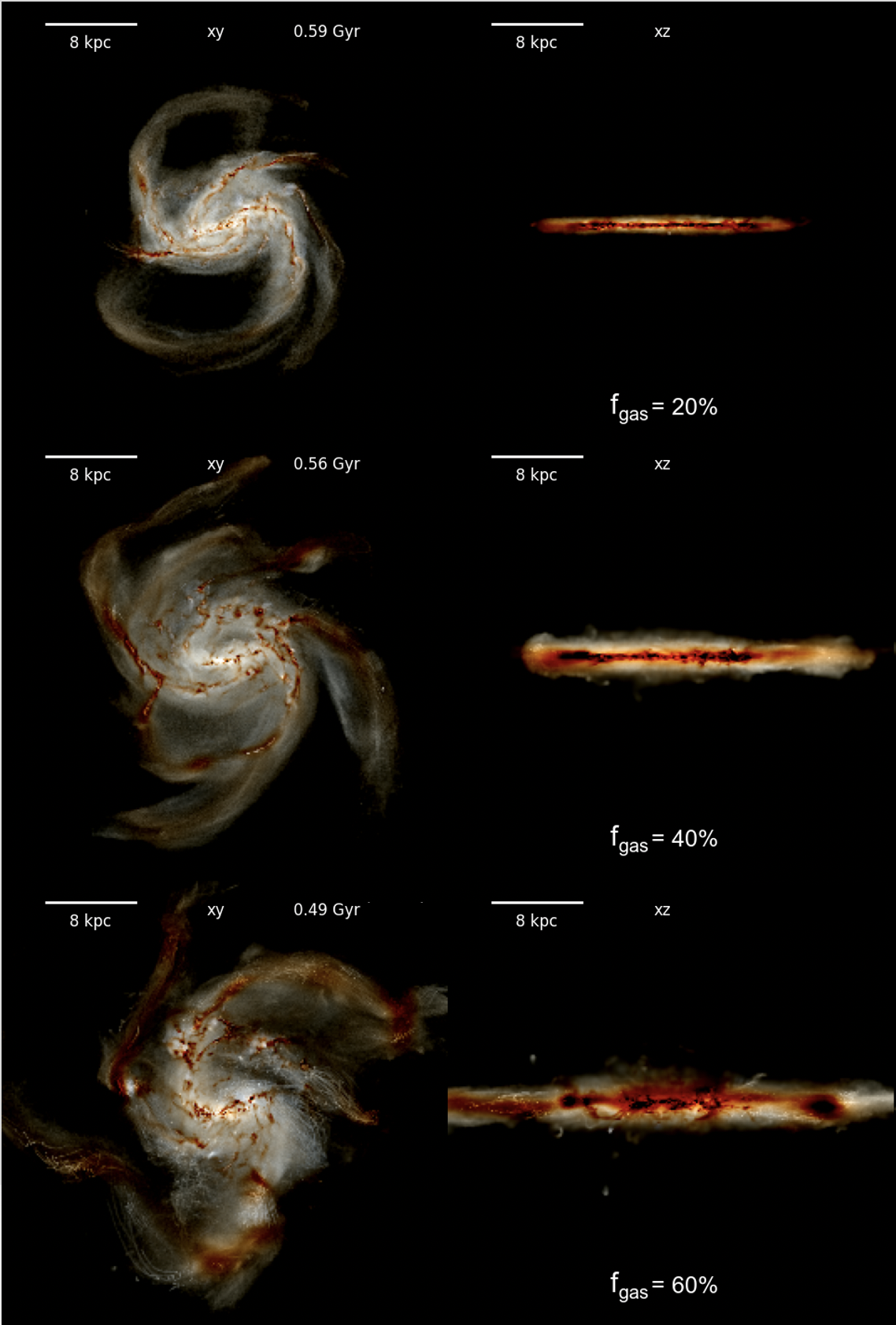}
    \caption{Three snapshots for three different values of $f_{\rm gas}$ (20\%, 40\%, 60\%) from top to bottom. The images are shown at roughly the time the bar emerges in each case and are a composite of the gas and newly formed stars. The false-colour rendering follows the method of \citet{pont13} described in Appendix A. The reddened regions are where most of the metal-enriched gas (and therefore dust) is expected.}
    \label{f:snaps}
\end{figure*}

In summary, within the uncertainties, we meet the photometric and size requirements for the three non-lensed $z\gtrsim 4$ bars listed in Table~\ref{t:bars}. The bars are formed rapidly and bear a morphological similarity to the observations (Fig.~\ref{f:snaps}).

\subsection{The bar's evolution after saturation}

As the stellar disc becomes increasingly dominant relative to the local dark matter, it tends to develop a stronger bar, particularly for low values of $\sigma_z/\sigma_R$. Such bars are susceptible to vertical bending modes (the firehose instability), which can substantially weaken or even destroy the bar \citep{com90,rah91}. An example of the destructive effect of strong buckling is our $(f_{\rm disc},f_{\rm gas})=(70\%,0\%)$ model in Fig.~\ref{f:lastbar2}.

A higher gas fraction, however, tends to suppress the effects of buckling \citep{ber07,deb06,woz09,ath13}. Previous studies considered relatively cold gas fractions of $f_{\rm gas}\lesssim20\%$. Two mechanisms are likely to contribute to this stabilisation: (i) bar-driven inflow of gas toward the centre, which increases the local vertical oscillation frequency; and (ii) viscous dissipation in the gas, which damps vertical motions and suppresses the growth of bending modes.
We extend these earlier results to substantially higher gas fractions, reaching $f_{\rm gas}=60\%$ for $f_{\rm disc}=70\%$, while \citet{bla24,bla25} explore the full range of $f_{\rm gas}$ for $f_{\rm disc}=50\%$.
The stabilising influence of gas on bar survival is evident in Fig.~\ref{f:lastbar2}. 

\begin{figure*}
    \includegraphics[width=0.685\columnwidth]{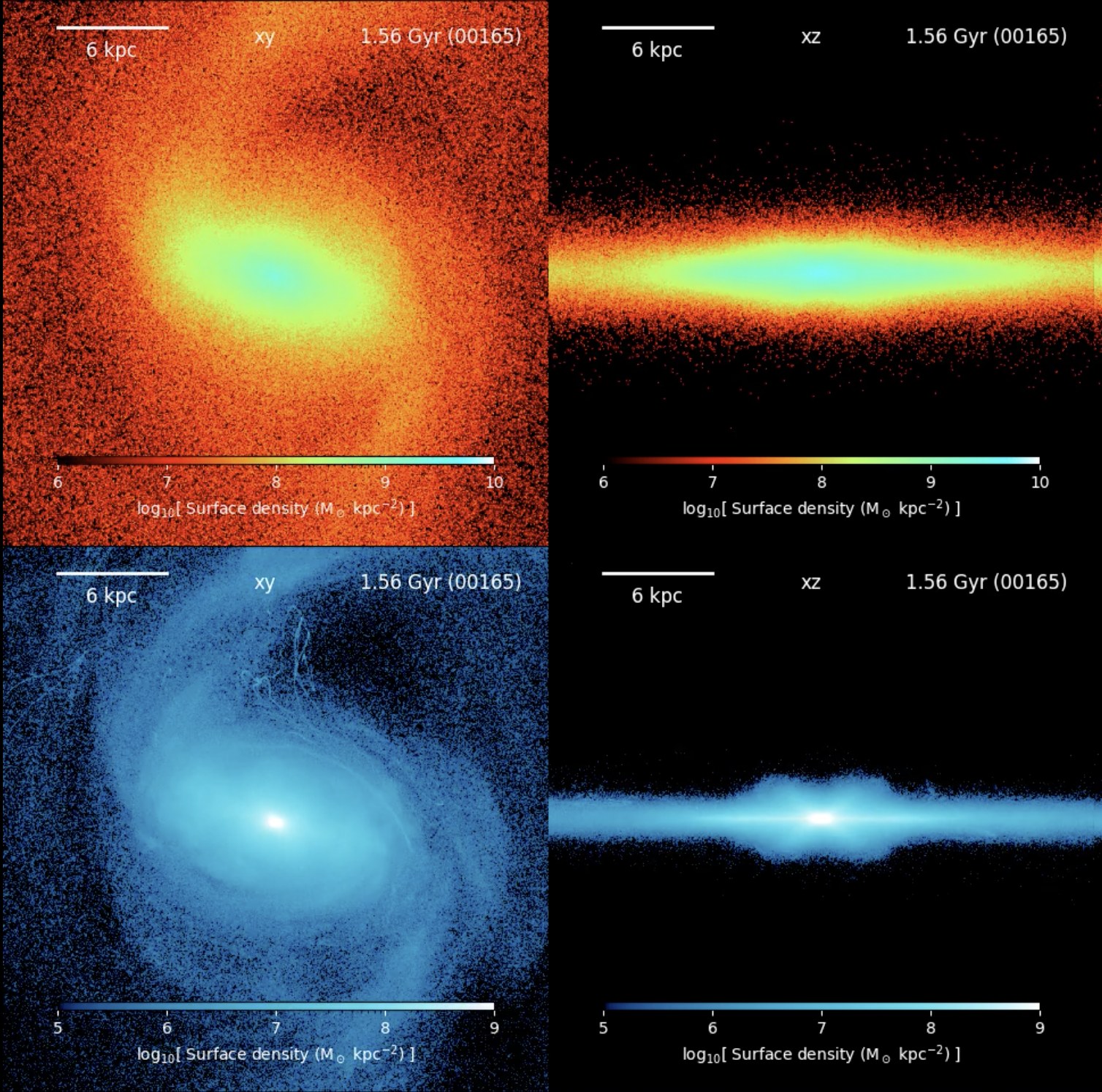}
    \includegraphics[width=0.685\columnwidth]{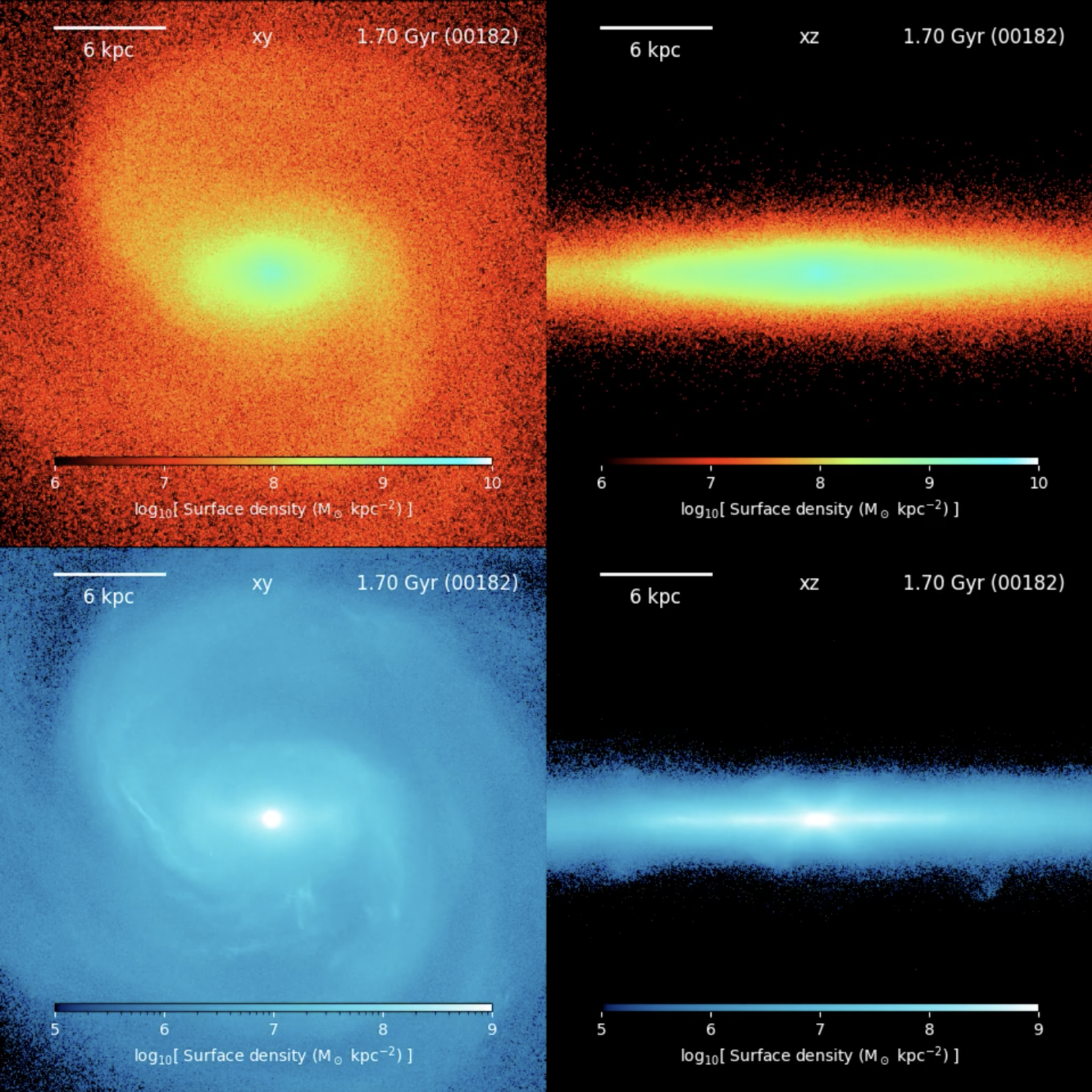}
    \includegraphics[width=0.685\columnwidth]{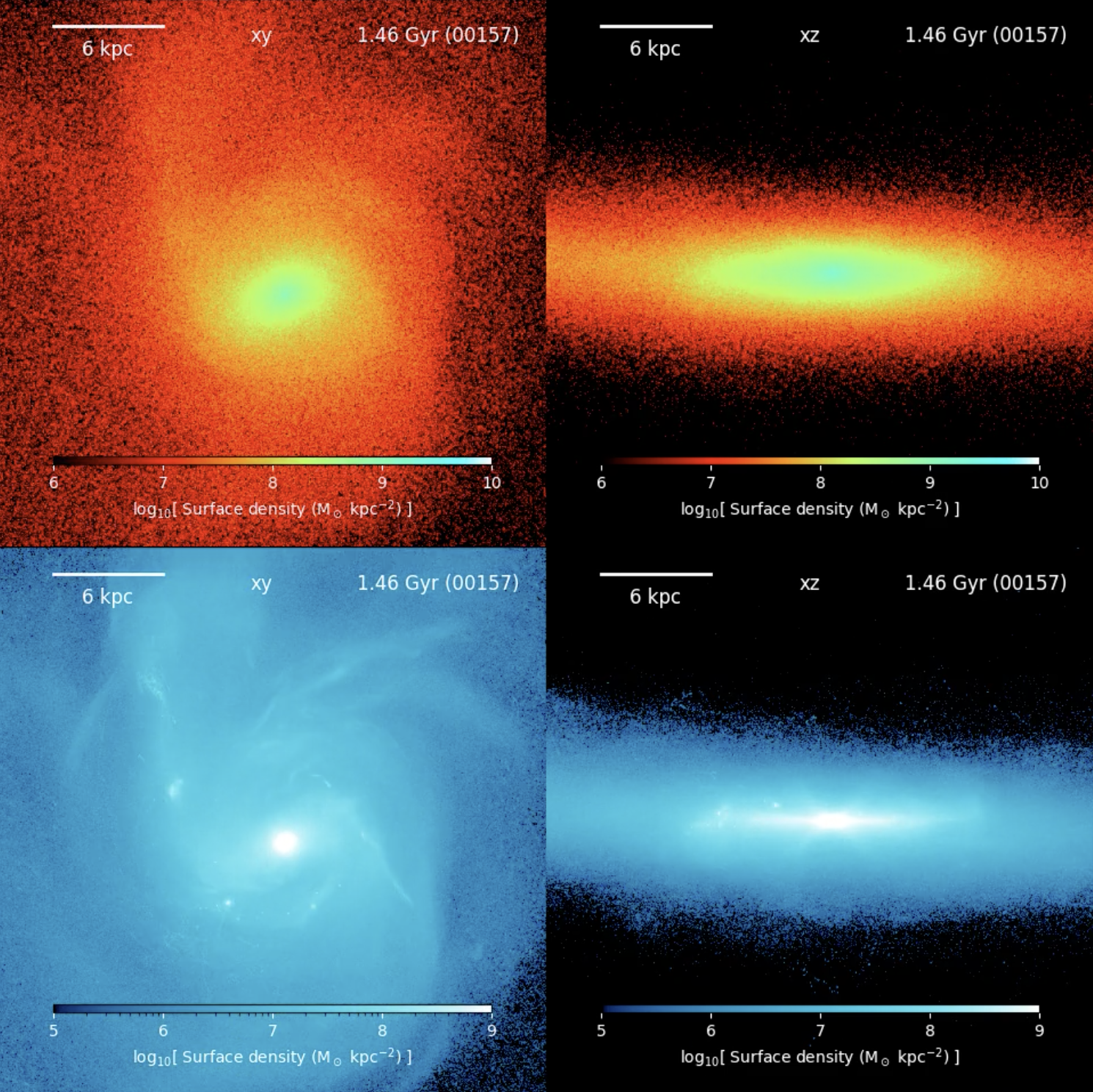}
    \caption{Three snapshots for three different values of $f_{\rm gas}$ (20\%, 40\%, 60\%) from left to right. Within each panel, the top row is the XY (left) and XZ (right) projections for the pre-existing (older) stars. The bottom row is the XY (left) and XZ (right) projections for the newly formed stars. In all cases, a box-peanut bulge is observed in the XZ projections; these arise at late times when the bar's initial peak strength has declined substantially. The strengths of the central bar and box-peanut bulge are inversely correlated with $f_{\rm gas}$.}
    \label{f:Xbar}
\end{figure*}

It is well established that, in gas-free discs, box-peanut (B/P) bulges typically develop after the bar achieves its peak amplitude \citep[q.v.][]{asano26}.  
But Fig.~\ref{f:Xbar} shows that, for the $f_{\rm disc}=70\%$ models, an X-shaped, box-peanut bulge develops in all gas-rich cases ($f_{\rm gas}=20,40,60\%$), appearing $200$--$400$ Myr after bar saturation during the subsequent settling phase. In contrast, the $f_{\rm gas}=0\%$ model undergoes strong buckling that substantially weakens the bar before a fully developed B/P structure can emerge.

The contrast of the X-structure decreases systematically with increasing $f_{\rm gas}$, becoming barely discernible at $f_{\rm gas}=60\%$, consistent with the progressive weakening of the bar. In all three gas-rich models, the X-structure remains aligned with and rotates with the bar, and is most prominent in side-on projection. Its orbital signature, however, becomes increasingly intermittent with increasing $f_{\rm gas}$, with strong intermittency extending across the bar even at late times. The delayed and intermittent appearance of the X-structure, together with its association with resonant three-dimensional orbit families, suggests that classical buckling may not be the primary mechanism populating the B/P structure in these gas-rich systems \citep[cf.][]{sel20}.

For fixed $f_{\rm gas}$, the X-shaped B/P structure is stronger in the $f_{\rm disc}=70\%$ models than in their $f_{\rm disc}=50\%$ counterparts \citep{bla24,bla25}. This behaviour is consistent with orbital theory: as the bar strengthens, stars trapped around the planar $x_1$ family encounter vertical resonances at which this family becomes unstable, giving rise to three-dimensional orbit families that support vertically extended B/P/X-shaped structures \citep[q.v.][]{ath05}.

\section{Summary remarks}

High-redshift stellar bars are remarkable because they demonstrate that massive, dynamically coherent discs were already assembled within the first Gyr of cosmic history. 
This work is in response to the recent discoveries (Table~\ref{t:bars}) of early massive stellar bars in gas-rich, disc-dominant galaxies beyond $z=4$ \citep{sma23,tsu24,boo25,wang26}. Motivated by the growing population of barred galaxies at $z\gtrsim4$, we have used high-resolution turbulent gas+star simulations to investigate how such bars can form rapidly and survive in gas-rich, disc-dominated systems.

Our principal result is that two conditions are required. First, the disc must be sufficiently self-gravitating, with $f_{\rm disc}\simeq70\%$, so that a global $m=2$ instability can grow on a timescale of a few hundred Myr. Secondly, a substantial gas fraction is required to suppress the vertical buckling that otherwise destroys a strong bar in such a dominant stellar disc. At $f_{\rm disc}=70\%$, bars form within 0.8 Gyr for $f_{\rm gas}\gtrsim20\%$, and within 0.4 Gyr for $f_{\rm gas}\simeq60\%$. The accreting and non-accreting models give essentially the same result, indicating that the rapid formation is not primarily driven by continued halo accretion. 

{\it We recognize that the $z\approx5$ barred spiral galaxy discovered by \citet{wang26} provides an especially stringent constraint on any scenario for the formation of massive discs and bars.} The galaxy is observed only 1.2 Gyr after the Big Bang, leaving little time for the assembly, settling and subsequent bar formation of a massive, dynamically cold disc. In our models, the required formation timescale depends sensitively on the gas fraction. At the higher gas fractions, where bars can develop most rapidly, the disc must nevertheless have been assembled and dynamically stabilised by $z\sim 7$. At lower gas fractions, the bar requires a longer period of coherent disc growth, pushing the formation and stabilisation of the disc back to $z\sim 12$. Thus, the observed $z\approx 5$ bar provides a stringent lower bound on the epoch at which a massive, rotationally supported disc must have emerged. In this sense, the bar is not simply evidence for an early stellar structure; it places a constraint on the entire preceding history of disc assembly and dynamical settling.

The dependence on gas fraction is systematic. Increasing $f_{\rm gas}$ produces earlier but weaker and smaller stellar bars, with the bar mass declining from $\sim3\times10^{10}$ M$_\odot$ at $f_{\rm gas}=20\%$ to $\sim1.5\times10^{10}$ M$_\odot$ at $f_{\rm gas}=60\%$. The corresponding bar sizes also decrease with increasing gas fraction, while the pattern speed shows only a weak systematic dependence.
We interpret this behaviour as the competition between two effects of turbulent gas. Gas fluctuations provide a stochastic forcing that increases the probability of excursions into the nonlinear regime where resonant orbit trapping can establish a bar, thereby reducing its onset time. At the same time, the same fluctuations disrupt phase coherence and reduce the efficiency of orbital trapping, limiting the final bar amplitude. A simple Langevin model incorporating the measured gas-density fluctuations reproduces the qualitative dependence of bar onset and saturation amplitude on $f_{\rm gas}$. The model should be regarded as a phenomenological description of this process.

The gas also fundamentally changes the subsequent evolution. In the $f_{\rm disc}=70\%$ models, gas suppresses catastrophic buckling, allowing a boxy-peanut/X-shaped structure to develop after bar saturation. These structures appear 200-400 Myr after saturation and persist even at $f_{\rm gas}=60\%$, although their contrast and orbital coherence decline systematically with increasing gas fraction.

Finally, the simulated $f_{\rm disc}=70\%$, $f_{\rm gas}=20-60\%$ systems reproduce, within the observational uncertainties, the masses, sizes and morphologies of the currently known non-lensed ($z\gtrsim4$) barred galaxies. The bars remain actively star-forming rather than quenched, providing a clear observational distinction between these early systems and the predominantly passive bars of the present-day Universe. Thus,
drawing on the new models, we predict that the high redshift, extreme discs should have actively star-forming bars, rather than merely stellar bars embedded in star-forming galaxies. A particularly testable prediction is that the earliest bars should be sites of vigorous star formation, with no strong suppression of star formation by bar shear \citep{bla24}.

Taken together, our results suggest two different pathways for the evolution of massive bars. At early times, rapidly growing, gas-rich and disc-dominated galaxies can produce massive stellar bars within a few hundred Myr. At later times, after gas depletion and secular evolution, bars can instead emerge and grow slowly over several Gyr. Early massive bars therefore need not be relics of long-term secular evolution: they can be a natural consequence of the extreme, turbulent and baryon-dominated conditions prevailing in young galaxies.

The results presented here point towards a broader change in the dynamical character of gas-rich stellar discs. As we show in our next paper (Villa-Alatorre et al 2026, in prep.), as $f_{\rm gas}$ increases, the stellar orbits become progressively more chaotic, signalling that the evolution of the disc cannot be understood simply as a scaled version of the familiar collisionless bar instability. We examine this transition in detail and show how the increasing importance of chaotic orbits provides a dynamical explanation for the weakened and increasingly intermittent structures found in the most gas-rich models. This extends the picture emerging from our earlier work \citep{bla24,bla25}: increasing gas fraction does not merely delay or suppress conventional secular evolution, but drives the stellar disc into a qualitatively different dynamical regime. Taken together, these results suggest that the evolution of early gas-rich galaxies is governed by a competition between coherent resonant structure and increasingly stochastic, chaotic stellar dynamics, a regime that has no direct analogue in present-day, gas-poor barred galaxies.
 

\section*{Acknowledgements}

We are grateful to Eugene Vasiliev and Romain Teyssier for continued assistance and adaptions with \agama\ and \ramses\ respectively. JBH is indebted to the University of Tokyo and to his hosts Professor Aihara and Dr Michiko Fujii during April 2026.
TTG acknowledges financial support from the Australian Research Council (ARC) through Australian Laureate Fellowships awarded to JBH (FL140100278) and TRB (FL220100117), for partial funding from Lund University, and from the James Arthur Pollock memorial fund awarded to the School of Physics, University of Sydney. JBH also acknowledges support from the ARC grant DP220103384 shared with Ken Freeman. OA acknowledges support from the Knut and Alice Wallenberg Foundation, the Swedish Research Council (grant 2019-04659) and the Swedish National Space Agency (SNSA Dnr 2023-00164). 

The computations and data storage were enabled by two facilities: (i) the National Computing Infrastructure (NCI) Adapter Scheme, provided by NCI Australia, an NCRIS capability supported by the Australian Government; and (ii) LUNARC, the Centre for Scientific and Technical Computing at Lund University (resource allocations LU 2023/2-39 and LU 2023/12-6). We further acknowledge high-performance computing resources provided by the Leibniz Rechenzentrum and the Gauss Centre for Supercomputing (grants~pr32lo, pr48pi, pn76ga and GCS Large-scale project~10391), the Australian National Computational Infrastructure (grant~ek9) and the Pawsey Supercomputing Centre (project~pawsey0810) in the framework of the National Computational Merit Allocation Scheme and the ANU Merit Allocation Scheme.

\section*{Data Availability}

The data underlying this article will be shared on reasonable request to the corresponding author.



\bibliographystyle{mnras}
\bibliography{main,federrath}


\appendix

\section{False-colour images}

Synthetic galaxy images were generated using the package developed by \citet{pont13}, which calculates the luminosities of star particles based on stellar population synthesis models. Each particle in the simulation represents a simple stellar population (SSP), corresponding to a coeval population of stars with a single age and metallicity. Magnitudes in photometric bands are obtained by interpolating pre-computed SSP tables—such as those from the Padova models (https://stev.oapd.inaf.it/cgi-bin/cmd) that provide the expected luminosity of a 1 M$_\odot$ SSP as a function of age and metallicity. These intrinsic magnitudes are then scaled according to the stellar mass of each particle, such that the luminosity of a particle is directly proportional to its mass. This ensures that the resulting photometry captures the dependence of colour and brightness on both age and chemical composition of the stellar population, with younger, more metal-poor populations appearing bluer and more luminous, and older, metal-rich populations appearing redder and fainter.

To construct false-colour images, the particle luminosities are projected onto a two-dimensional grid, producing surface brightness maps in multiple photometric bands. Magnitudes are converted to linear fluxes, which are then mapped to RGB channels to create composite images. The choice of bands for the RGB channels is flexible, allowing the visualisation to highlight specific stellar population properties or spectral features. The resulting images encode relative stellar population information rather than human-visible colours.

Extinction by interstellar dust is incorporated using a simplified foreground screen model. The optical depth along each line of sight is estimated from the gas column density and metallicity, assuming a constant dust-to-metal ratio. Extinction in each band is then calculated using a standard extinction law (i.e. Calzetti law with $R_V = 3.1$), and applied to the surface brightness maps. This produces reddening and attenuation of the light from star particles, approximating the effect of dust on the observed photometry without performing full radiative-transfer calculations. To account for depth effects in the disc, the model typically assumes half of the dust is in front of and half behind the stars along each line of sight.

This combination of SSP-based photometry and first-order dust extinction modelling allows for the generation of physically motivated synthetic galaxy images. The approach preserves the key stellar population characteristics of the simulated galaxies while providing a visually informative representation of how age, metallicity, and dust influence the integrated light, enabling direct comparison with observational data and analyses of galaxy morphology, colour gradients, and dust effects.

\bsp	
\label{lastpage}
\end{document}